\documentclass[fleqn,usenatbib]{mnras}
\usepackage{amsmath}
\usepackage{graphics}
\usepackage{color}

\usepackage{newtxtext,newtxmath}

\usepackage[T1]{fontenc}

\DeclareRobustCommand{\VAN}[3]{#2}
\let\VANthebibliography\thebibliography
\def\thebibliography{\DeclareRobustCommand{\VAN}[3]{##3}\VANthebibliography}

\usepackage{graphicx}	
\usepackage{amsmath}	

\title[Frequency Shift in Binary Microlensing]{Frequency Shift in Binary Microlensing}

\author[Samaneh Sarbaz and Sohrab Rahvar]{
Samaneh Sarbaz,$^{1}$\thanks{E-mail:samaneh\_sarbaz@physics.sharif.edu}
Sohrab Rahvar$^{{1},{2}}$\thanks{E-mail:rahvar@sharif.edu}
\\
$^{1}$Department of Physics, Sharif University of Technology , Azadi ave.
Tehran 11365-9161, Iran \\
$^{2}$Research Center for High Energy physics , Sharif university of Technology , Tehran , Iran \\
}

\date{Accepted XXX. Received YYY; in original form ZZZ}

\pubyear{2024}

\begin{document}
\label{firstpage}
\pagerange{\pageref{firstpage}--\pageref{lastpage}}
\maketitle

\begin{abstract}
Gravitational microlensing with binary lensing is one of the channels for detecting exoplanets. Due to the degeneracy of the lens parameters for the binary microlensing, additional features such as parallax and finite-size effects need to identify the lens parameters. The frequency-shift effect as the relativistic analogy of the gravity assist for the photons, is an extra observation that provides additional constraint between the lens parameters .  
In this work, we extend the application of the frequency shift effect to binary microlensing and  
derive the frequency shift during the lensing and caustic crossing. The frequency shift for the binary lens is of the order of    ${\Delta\nu}/{\nu}\sim 10^{-12}$.
We also investigate the feasibility of detecting this effect by employing Cross-Correlation methods .
\end{abstract}




\section{Introduction} \label{sec:1}
Gravitational lensing is one of the predictions of General Relativity that confirmed by Arthur Eddington in the 1919 eclipse. The gravitational lensing of a star by another compact object has been studied by  \cite{pac86} and observed for the first time by \cite{SupernovaCosmologyProject:1993faz} . In the so-called gravitational microlensing,  the temporal magnification of the background star is observable. 
The duration of the lensing is called Einstein-crossing time (i.e. $t_E$) and inside the Galaxy, it is of the order of the month. The Einstein crossing time is a function of the kinematics of the lens, source, distance, and mass of the lens. This makes the problem degenerate with respect to the lens parameters. On the other hand, the perturbation effects in the light curve such as the parallax \citep{Gould_1998,rahvar2003,Novati_2015,Street_2016,Bachelet_2012,Shvartzvald_2019,Hirao_2020,Zang_2020,Shvartzvald_2016} and finite-size effect \citep{witt,Alcock_1997} make a small deviation of the light curve from simple lensing. These perturbation effects can break the degeneracy and provide a unique solution for the microlensing event, especially for the detection of exoplanets orbiting around the lens stars \citep{Rahvar_2015}. The statistical analysis of gravitational microlensing, integrating over the degenerated parameters also has been used for studying the structure of the Galaxy \citep{2009A&A...500.1027R,Moniez_2017}.

The other method for breaking the degeneracy between the lens parameters is the spectroscopy of microlensing source star, using the gravity assist effect or so-called Shapiro effect for photons \citep{Rahvar_2020}.
In this effect, the positive and negative parity images from the microlensing gain and lose energy depending on the transverse velocity of the lens with respect to the position of images.  The result is a time-dependent frequency shift of the source star. The observation of this effect provides extra information about the transverse velocity of the lens.

In this work, we extend the frequency shift in microlensing to the binary lensing systems which is important from the point of view of the discovery of the exoplanets \citep{Gould_2010,2011A&A...529A.102B,Muraki_2011,Sahu_2022,Choi_2013,Street_2013,2013A&A...552A..70K,Shin_2012,Skowron_2015}. Nowadays gravitational Microlensing is one of the standard methods for the discovery of exoplanets besides the other methods such as the transit, Doppler effect, and astrometry of the stars. In binary lensing, we have two extra parameters of $\chi$ representing the distance of the two binary lenses and $\mu$ as the mass ratio of the lenses. In general, there is degeneracy between the solution for the light curve of binary lenses and we have no unique solution for the light curves. However, the perturbation effects as well as the frequency shift measurements can break partially the degeneracy between the lens parameters. The detection of binary microlensing events is not rare. For instance, the OGLE project which is observing the microlensing events along the center of the Galaxy reports the list of microlensing events \citep{Mr_z_2020} and out of $170$ gravitational microlensing events, about $25$ of them are binary which means that $14\%$ of microlensing events in this list. Some of the binary microlensing events contains the caustic crossing \footnote{ In which the source passes through caustic lines is called caustic crossing} in the light curves which causes a sharp peak in the light curves.

In this work, we investigate the effect of frequency shift in the binary lenses and the observational signature for breaking the degeneracy. The organization of this paper is as follows: In Section (\ref{2}) we review the binary lensing. Section (\ref{3}) discusses the frequency shift effect in single lensing and binary lensing. We will investigate and discuss the feasibility of the frequency shift measurements by the present and future spectrographs in section (\ref{subsec:3-5}). The conclusion is given in section (\ref{4}). 
\section{Lens equation for Microlensing by binary lenses}\label{2} \label{subsec:2-2}
\begin{figure}
\begin{center}
\includegraphics[width=\columnwidth]{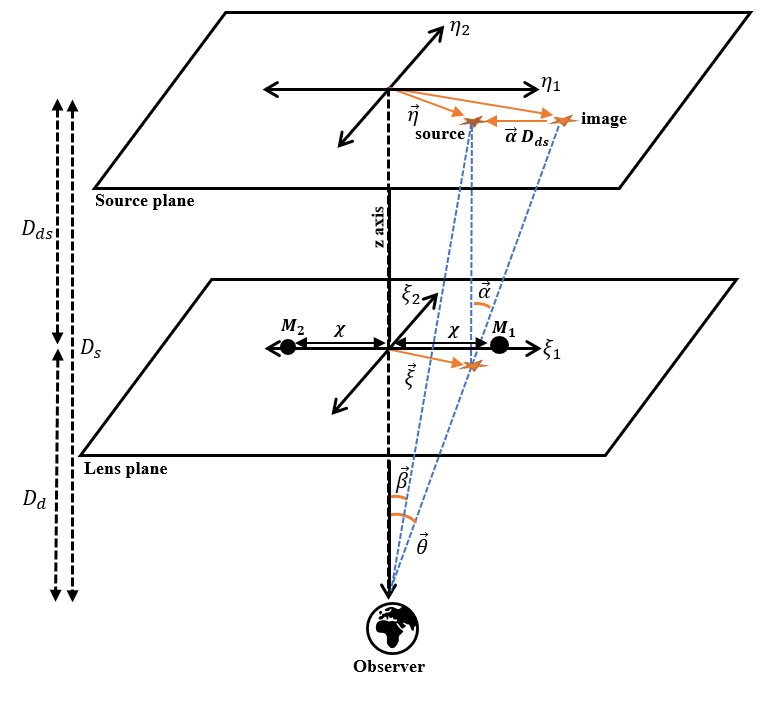}
\end{center}
\caption{Overview of the location of the observer and  binary lenses in their coordinate system
\label{fig:3}}
\end{figure}

 A brief review to calculate the lens equation for Microlensing by single lens is available in \ref{subsec:2-1} where we adapt the notation in
\cite{Rahvar_2020}.  To get started, we define three coordinate systems. One of them is the source Cartesian coordinate system with the axis of $\eta_1 $ and $\eta_2$, the second one is the lens coordinate system with the axis of $\xi_1$  and $\xi_2$. The third coordinate system is the observer coordinate system. In addition, we assume that both the source and lens are point-like masses (see Figure \ref{fig:3}).\\
For binary lensing system , we review the gravitational microlensing by binary lenses with the masses of $ M_1 $ and $ M_2 $  \citep{1986A&A...164..237S, 1999A&A...349..108D, 1993A&A...268..453E}.  We set the center of the coordinate on the lens plane at the point which is exactly in the middle of the distance between the two lenses (see Figure \ref{fig:3}) and define the coordinate system where two lenses are located on $\xi_1$ axis at equal distances from the origin. The vectors representing the position of the two lenses are $\vec{\rho_1}$ and $\vec{\rho_2}$.It should be noted that in this section, following \citep{1986A&A...164..237S} , we will work with the spatial vectors, not the angle vectors that were used in the single lens.
 
 In the source plane, we define a coordinate system  $(\eta_1,\eta_2 )$ which is oriented parallel to the $(\xi_1,\xi_2 )$ and the center of the coordinate system is on the point of intersection with the optical axis. Then the light bending is given by 
 
\begin{equation}\label{equ:20}
   \vec{\alpha}(\vec{\xi})=\frac{4GM_1}{c^2}\frac{\vec{\xi}-\vec{\rho}_1}{|\vec{\xi}-\vec{\rho}_1|^2}+\frac{4GM_2}{c^2}\frac{\vec{\xi}-\vec{\rho}_2}{|\vec{\xi}-\vec{\rho}_2|^2}, 
\end{equation}
where the lens equation is
\begin{equation}\label{equ:22}
\vec{\eta}=\frac{\vec{\xi}}{D_d}D_s-\vec{\alpha}(\vec{\xi})D_{ds}
\end{equation}
and the Einstein radius according to the definition of Einstein angle in equation (\ref{equ:11}) is :
\begin{equation}\label{equ:23}
 \rho_E=\sqrt{\frac{4GMD_d \ D_{ds}}{c^2D_s}}
\end{equation}
here for binary lenses, $M = M_1 + M_2$. We define two new parameters as
\begin{equation}\label{equ:24}
 \vec{r}=\frac{\vec{\xi}}{\rho_E},
\end{equation}
\begin{equation}\label{equ:25}
 \vec{x}=\frac{D_d}{D_s}\frac{\vec{\eta}}{\rho_E},
\end{equation}
where $ \vec{r}=(r_1,r_2) $ is the normal vector of the lens coordinate system and shows the position of the images on the lens plane after the deflection by the lenses, and $ \vec{x}=(x_1,x_2) $ is the normal vector that indicates the position of the source. Also for simplicity, we define:

\begin{equation}\label{equ:26}
\mu_1=\frac{M_1}{M} \ \ , \ \ \mu_2=\frac{M_2}{M}
\end{equation}

Therefore, after substituting, the lens equation simplifies to:
\begin{equation}\label{equ:28}
\vec{x}=\vec{r}-\vec{\underline\alpha}(\vec{r}),
\end{equation}
where the $\vec{\underline\alpha}(\vec{r})$ is the deflection angle normalized to the Einstein radius simplified as follows:
\begin{equation}\label{equ:29}
\vec{\underline\alpha}(\vec{r})=\mu_1\frac{\vec{r}-\vec{r_{(1)}}}{|\vec{r}-\vec{r_{(1)}}|^2}+\mu_2\frac{\vec{r}-\vec{r_{(2)}}}{|\vec{r}-\vec{r_{(2)}}|^2},
\end{equation}
where the lenses are located at 
$\vec{r}_{(1)}=-\vec{r}_{(2)}=\chi\hat{\xi}_1$.
\\
The magnification is obtained from $A=\left|\frac{\partial x}{\partial r}\right|^{-1}= \left|\frac{1}{J} \right| $ according to the Jacobian as follows
\begin{equation}\label{equ:37}
J=\frac{\partial\vec{x}}{\partial\vec{r}}=\left(\begin{matrix}
\frac{\partial{x}_1}{\partial{r}_1} & \frac{\partial{x}_1}{\partial{r}_2} \\
\frac{\partial{x}_2}{\partial{r}_1} & \frac{\partial{x}_2}{\partial{r}_2}
\end{matrix}\right).
\end{equation}

\begin{figure}
\begin{center}
\includegraphics[width=\columnwidth]{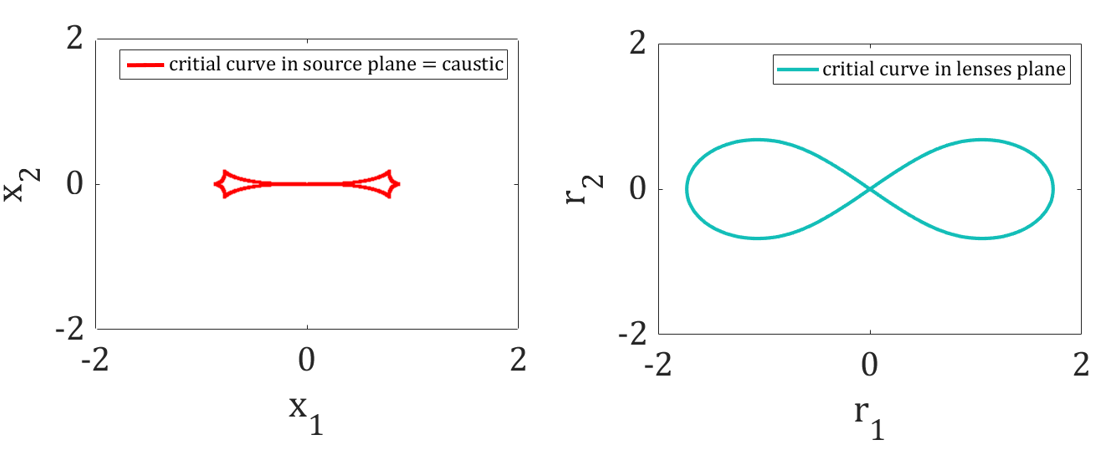}
\end{center}
\caption{Critical curve for symmetric binary lenses $\mu_2=\mu_1=0.5$ and $\chi=1$
\label{fig:4}}
\end{figure}

\begin{figure}
\begin{center}
\includegraphics[width=\columnwidth]{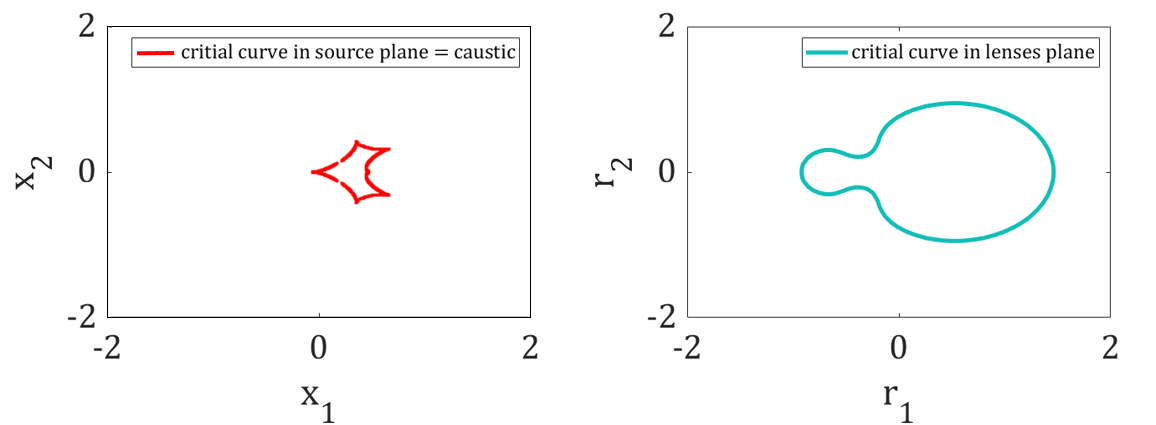}
\end{center}
\caption{Critical curve for asymmetric binary lenses $\mu_2=0.1 \ \mu_1=0.9 $ and $\chi=0.5$
\label{fig:5}}
\end{figure}

\begin{figure}
\begin{center}
\includegraphics[width=\columnwidth]{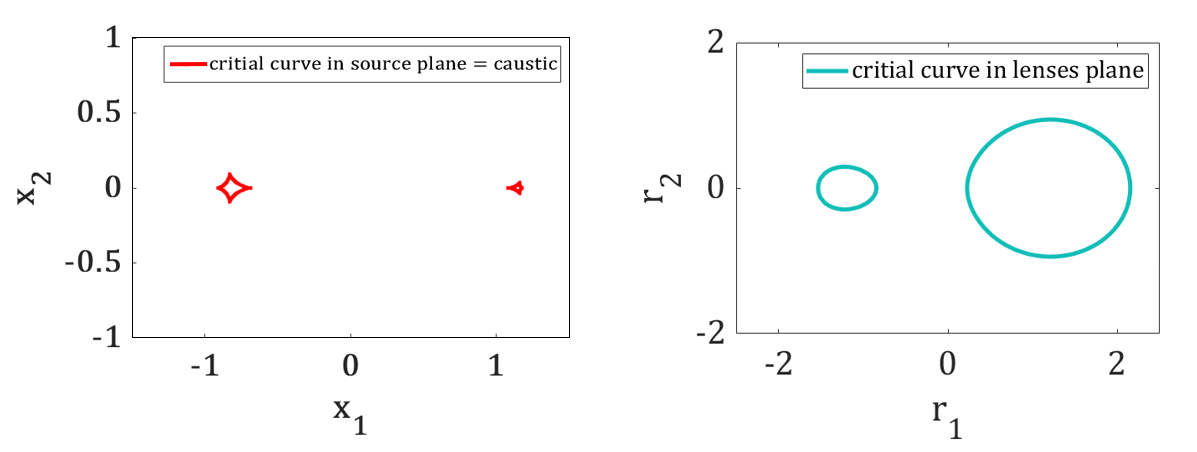}
\end{center}
\caption{Critical curve for asymmetric binary lenses  $\mu_2=0.1$ and  $\mu_1=0.9$ and $\chi=1.2$
\label{fig:305}}
\end{figure}

One of the important features of binary lensing is the critical curves where for certain values of $\vec{r}=(r_1,r_2)$, the determinant of the matrix J is zero (i.e. $\textrm{det}J=0.
$).
Substituting $x_i$ in terms of $r_i$ results in the critical line as \citep{1999A&A...349..108D}
\begin{align}\label{equ:40}
 &\left((r_1-\chi)^2  + (r_2)^2\right)^2 \left((r_1+\chi)^2+(r_2)^2\right)^2\notag\\
  -& \mu_1^2 \left((r_1+\chi)^2+(r_2)^2\right)^2 -\mu_2^2\left((r_1-\chi)^2+(r_2)^2\right)^2\notag\\
     - & 2\mu_1\mu_2\left((r_1-\chi)^2+(r_2)^2\right)^2    \left((r_1+\chi)^2+(r_2)^2\right)^2\notag\\
 +& 16\mu_1\mu_2 \chi^2 (r_2)^2 \ = \  0, 
\end{align}

where the corresponding curve on the source plane is called the caustic line. If a point-like source crosses this curve we will have infinite magnification however taking into account the wave optics a diffraction pattern is expected and we can avoid singularity \citep{Nakamura,mehrabi}. There are three different topologies of critical curves and caustics of a binary lenses \citep{1999A&A...349..108D}. Samples of the critical curve and caustic lines are given in Figures (\ref{fig:4}) for the parameters of $\mu_2=\mu_1=0.5$ and $\chi=1$ and for the parameters of $\mu_2=0.1 \ \mu_1=0.9 $ and $\chi=0.5$ in Figure (\ref{fig:5}). Also Figure (\ref{fig:305}) is critical curve and caustic of binary lenses with the parameters of $\mu_2=0.1$ and  $\mu_1=0.9$ and $\chi=1.2$

\begin{figure}
\begin{center}
\includegraphics[width=\columnwidth]{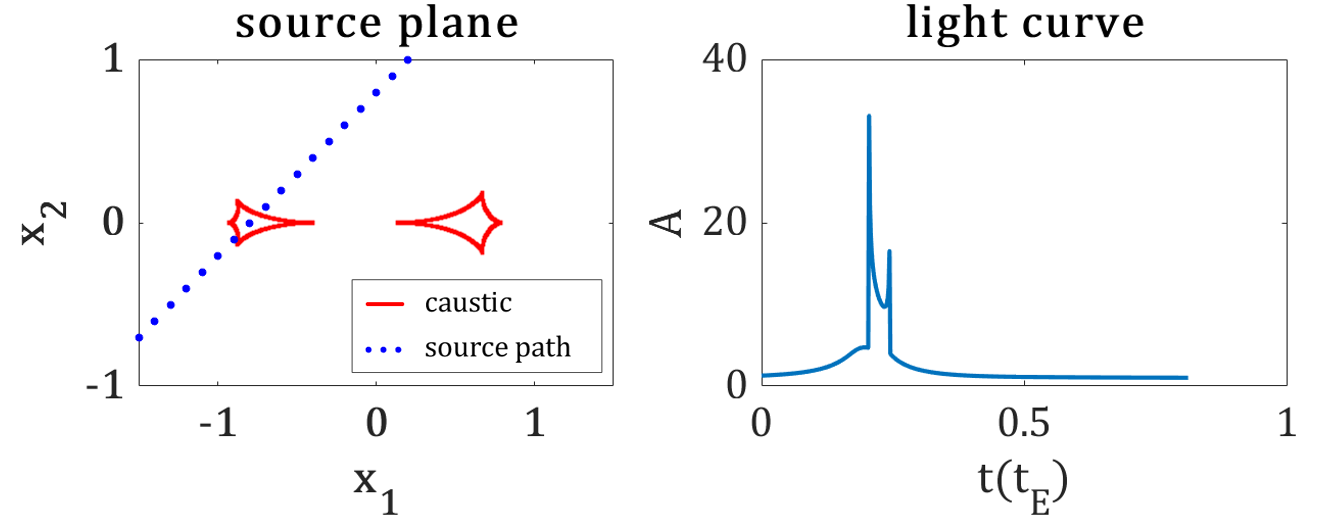}
\end{center}
\caption{The caustic curve and the motion of the source in the source's normalized coordinate system $ \vec{x}=\frac{D_d}{D_s} \frac{\eta}{\rho_E}$.
In this case, two lenses with mass ratios  $\mu_2=0.3$ and $\mu_1=0.7  $    each located at a distance $\chi = 1$  from the origin. The trajectory has an inclination angle of $45^\circ$ with respect to the horizontal axis with the minimum impact parameter of $0.57$.}
\label{fig:7}
\end{figure}

The number of images depending on the position of the source with respect to the caustic lines is either $3$ (outside) or $5$ (inside). For simplicity in the solution of the lens equation, we use label for identifying the image number as ($r_{1,i},r_{2,i}$) where $i$ goes from $(1-3)$ or $(1-5)$. Figure (\ref{fig:7}) shows the geometry of the trajectory of the source and corresponding light curve.
we note that we avoid singularities in the light curves during source star crossing the caustic lines , the technical reason is that the trajectory of source star in our code is made of descrete points.

\section{Frequency shift in the Microlensing} \label{3}
In this section we discuss the frequency shift phenomenon which has been introduced in \cite{Rahvar_2020}(see appendix \ref{app:1}).  Our aim is to extend this idea to binary lensing. In the single lensing, we have the Einstein crossing time ($t_E$) as the observable parameter which includes all the parameters of lensing as follows:
\begin{align} \label{equ:1}
  t_E = & 45.6 \textrm{ day} \left( \frac{D_s}{8.5 \textrm{ kpc}}\right)^{\frac{1}{2}}\left( \frac{M}{0.5 M_{\odot}}\right)^{\frac{1}{2}}\left(x\left(1-x\right)\right)^{\frac{1}{2}}&\notag\\
   & \times \left(\left|V_{E,{\bot}}-V_{s,{\bot}}-\frac{1}{x}(V_{E,{\bot}}-V_{L,{\bot}})\right|\frac{1}{220 \ km/s}\right)^{-1}&  
\end{align}
For the distance of the source in the direction of the center of the galaxy, we set $ D_s \sim 8.5 kpc $.
$M$  is the mass of lenses, $x ={D_l}/{D_s} $   is the ratio of the lens distance to the source distance from the observer, $  V_{E,\bot}$, $ V_{S,\bot} $ and $ V_{L,\bot}$ are the transverse relative velocities of the Earth, source and the lens in the Galactic frame \citep{mollerach2002gravitational}. In equation (\ref{equ:1})  by measuring the $t_E$, we cannot derive the mass, the distance, and the transverse velocities of the lens and source. This is the so-called degeneracy problem in gravitational microlensing. In order to break the degeneracy between the lensing parameters, the perturbation effects such as parallax and finite-size effects deviate the simple microlensing light curve and provide the possibility to break partially this degeneracy \citep{Rahvar_2015}. In this section, we review the frequency shift as a new method that provides extra physical information both in the single and binary lensing systems.
\subsection{ Frequency shift in single lens}\label{subsec:3-2}
Frequency shift in microlensing can be shown with two different approaches.The first one is presented in \citep{Rahvar_2020} based on special relativity and taking into account the deflection angle by lensing in general relativity which is explained in the appendix \ref{app:1} . The other equivalent approach is Shapiro effect, which again leads to the same results \citep{afshordi} .\\
\begin{figure}
\begin{center}
\includegraphics[width=\columnwidth]{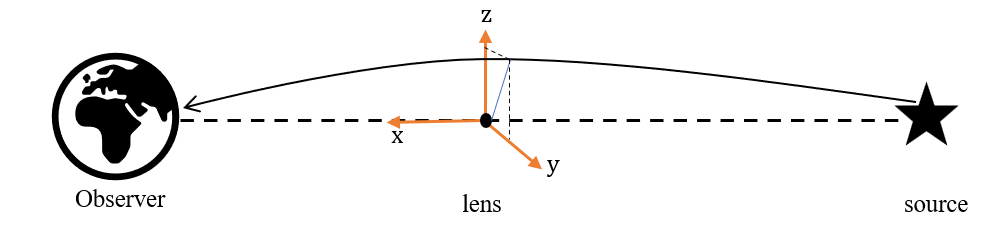}
\end{center}
\caption{for Shapiro effect we assumed the light emission  along the x-axis, and  the origin of coordinates on the lens on the x-axis
\label{fig:24}}
\end{figure}
Considering the lens as a point mass, we can start from the following metric
\begin{equation}\label{equ:88}
ds^2=-(1+2\Phi)dt^2+(1-2\Phi)dl^2
\end{equation}
so for null path $ds=0$,then
\begin{equation}\label{equ:89}
dt^2=(1-4\Phi)dl^2 \\ \rightarrow  t = \int (1-2\Phi)dl
\end{equation}
$l$ is in direction of our line of sight and $ \delta t =-2\int \Phi dl $ representing the time delay compare to flat space. So We can calculate the changes of $\delta t$ with respect to time as the result of gravitational potential :
\begin{equation}\label{equ:90}
\dot{\delta t} = -2\int (\frac{d\Phi}{dt})dl
\end{equation}
Since we know that $\dot{\delta t}=-\frac{\delta \nu}{\nu}$
\begin{equation}\label{equ:91}
\frac{\delta \nu}{\nu}= 2\int (\frac{\partial\Phi}{\partial t}+\Vec{v}.\nabla\Phi)dl
\end{equation} 
Where $\Vec{v}$ can be considered as the velocity of the lens relative to source. Considering the Newtonian potential for the single lens ,
\begin{equation}\label{equ:92}
\Vec{v}.\nabla \Phi = \left( \frac{GM} {r^3}(xv_x+yv_y+zv_z) \right)
\end{equation} 
Where $M$ is the lens mass , $G$ is gravity constant Let's assume that the direction of light emission is along the x-axis, and consider the origin of coordinates on the x-axis.(see figure \ref{fig:24}). If the gravitational potential of the lens does not change with time and only moves with respect to our line of sight, then :

\begin{equation}\label{equ:93}
\frac{\delta \nu}{\nu}= +2GM\int \left(\frac{xv_x+yv_y+zv_z} {(x^2+y^2+z^2)^{\frac{3}{2}}} \right) dx
\end{equation} 
Taking a constant velocity for the lens , the zero term for this integral is  the$\int^{\infty} _{-\infty} \left(\frac{xv_x} {(x^2+y^2+z^2)^{\frac{3}{2}}} \right) dx$ , as the odd function of x. 
Therefore, it seems that the frequency shift is independent of the lens speed along the line of sight.The impact parameter is a vector from the lens towards the image and we represent it by $\Vec{b} $ .By defining $b^2=y^2+z^2$ and $\Vec{b}.\Vec{v}=yv_y+zv_z $ the frequency shift simplifies to
\begin{align}\label{equ:94}
\frac{\delta \nu}{\nu}= 2GM \int^{\infty} _{-\infty}& \left(\frac{\Vec{b}.\Vec{v}} {(b^2+x^2)^{\frac{3}{2}}} \right) dx  = \notag \\
&2GM\Vec{b}.\Vec{V}\int^{\infty} _{-\infty} \left(\frac{dx} {(b^2+x^2)^{\frac{3}{2}}} \right) 
\end{align}
By defining $x=b \tan \theta $, the above result of integration is
\begin{equation}\label{equ:95}
\frac{\delta \nu}{\nu}= 2GM \frac{b}{b^3}\int^{\frac{\pi}{2}} _{\frac{\pi}{2}}\left(\cos{\theta} \right) d\theta = \frac{4GM}{b} \hat{b}.\Vec{v}
\end{equation} 
where in terms of deflection angle, that is
\begin{equation}\label{equ:96}
\frac{\delta \nu}{\nu}= |\alpha| \hat{b}.\Vec{v}
\end{equation} 
Here b is the location of the image relative to the lens and only the transverse velocity affects the frequency shift (see Figure \ref{fig:025}) 
\begin{figure}
\begin{center}
\includegraphics[width=\columnwidth]{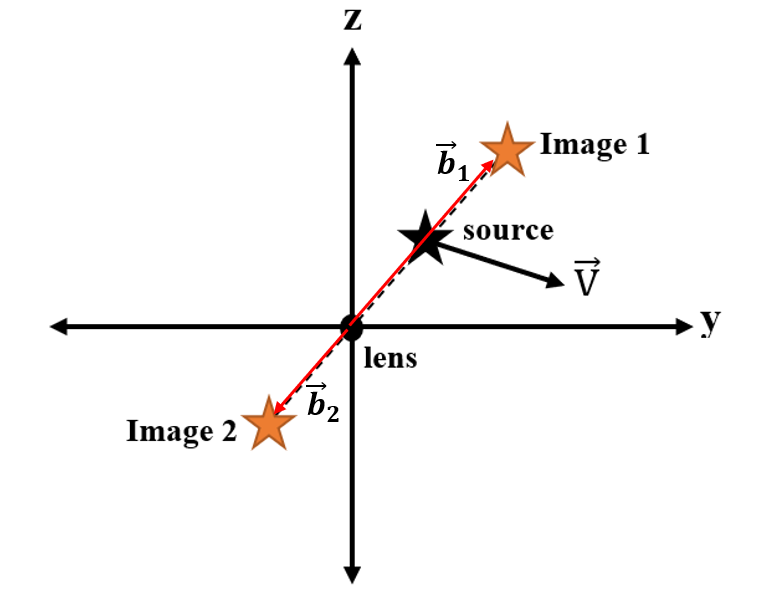}
\end{center}
\caption{Lens plane .As you could see $ \Vec{b}.\Vec{v} > 0$ is for image 1 and $ \Vec{b}.\Vec{v} < 0$ is for image 2.
\label{fig:025}}
\end{figure}
\subsection{Frequency shift in binary lenses}\label{subsec:3-3}
\begin{figure}
\begin{center}
\includegraphics[width=\columnwidth]{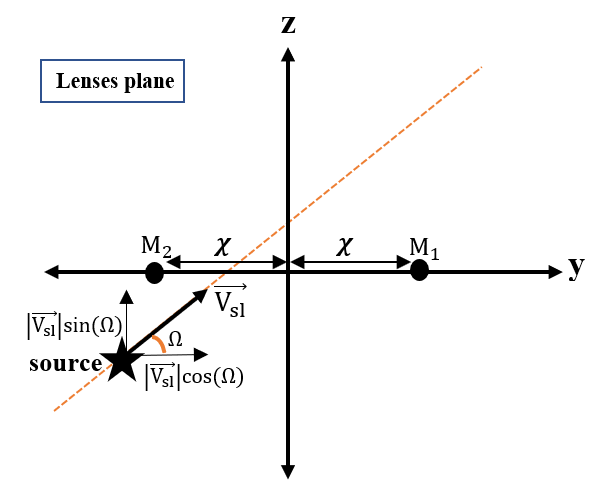}
\end{center}
\caption{ The relative velocity of source and lens in the lens coordinate system. The velocity vector has $\Omega$ angle with respect to the $\xi_1$ axis.  \label{fig:11}}
\end{figure}

Let us assume a source moving with a relative velocity compared to the center of mass of a binary lens (see Figure \ref{fig:11}). Here we use the Lorentz transformation and the Lorentz inverse transformation between the source and lens coordinates before and after the lensing, to find the relative energy change between the photons of the images. 
Starting with Figure (\ref{fig:3}), let us assume photons are emitted from the source along the z-direction. Therefore, the four-momentum of the photon in the source's coordinate system can be written as follows:

\begin{equation}\label{equ:41}
p^\mu =\left(\begin{matrix}
        p_z \\
        0 \\
        0 \\
        p_z
      \end{matrix}\right),
\end{equation}

The components of the relative transverse velocity of the source on the lens plane is 
\begin{equation}\label{equ:61}
\vec{V}_{sl} = \vec{V}_{sl}\cos\Omega \ \hat{\eta_1}+ \vec{V}_{sl}\sin\Omega \ \hat{\eta_2},
\end{equation}
where $\Omega$ is the angle of the source trajectory with respect to the $\hat{\eta_1}$ axis. For simplicity in calculation
we define the following parameters of
\begin{equation}
\label{equ:62}
\beta=\frac{\vec{V}_{ls}}{c} \quad
 \beta_x =\beta\cos\Omega \ , \quad \beta_y = \beta\sin\Omega    \end{equation}
Where $V_{ls}$ is the transverse velocity of the lens relative to the source. This transverse velocity can composed of stellar motion in galaxy and the rotation velocity of the binary lenses . Detail is given in Appendix \ref{app:3}\\
The Lorentz transformation between the source and lens plane is 
\begin{equation}\label{equ:63}
\Lambda = \left(\begin{matrix}
          \gamma & -\beta_x \gamma & -\beta_y\gamma & 0 \\
           -\beta_x\gamma & 1+\frac{(\gamma-1)\beta_x^2}{\beta^2} & \frac{(\gamma-1)\beta_x \beta_y}{\beta^2} & 0 \\
          -\beta_y\gamma & \frac{(\gamma-1)\beta_x \beta_y}{\beta^2} & 1+\frac{(\gamma-1)\beta_y^2}{\beta^2} & 0 \\
          0 & 0 & 0 & 1
        \end{matrix}\right).
\end{equation}
We apply the Lorentz transformation to the four-momentum components of photons which provides corresponding momenta in the lens plane.
\begin{equation}\label{equ:64}
\acute{p^\mu}=\left(\begin{matrix}
        \gamma p_z \\
        -\beta_x \gamma p_z  \\
        -\beta_y \gamma p_z  \\
        p_z
      \end{matrix}\right).
\end{equation}
The four-momentum of photons due to gravitational lensing gets extra terms because of deflection angle where for the binary lens is given by:
\begin{equation}\label{equ:65}
  \vec{\alpha}(\vec{\xi}=\rho_E \vec{r})=\frac{ D_s}{D_{ds}}\left[ \mu_1\frac{\vec{r}-\vec{r}_{(1)}}{|\vec{r}-{\vec{r}_{(1)}}|^2}+ \mu_2\frac{\vec{r}-\vec{r}_{(2)}}{|\vec{r}-{\vec{r}_{(2)}}|^2}\right]\theta_E,
\end{equation}
in terms of components of lens plane's bases vectors  ,the deflection angle can be written as 
\begin{align}\label{equ:66}
   \vec{\alpha}(\rho_E \vec{r})& =\frac{ D_s}{D_{ds}} \left(\frac{\mu_1 \ (r_1-\chi)}{(r_1-\chi)^2+(r_2)^2} + \frac{\mu_2 \ (r_1+\chi)}{(r_1+\chi)^2+(r_2)^2}\right)\theta_E\hat{\xi_1} \notag \\
   & +\frac{ D_s}{D_{ds}} \left(\frac{\mu_1 \ r_2}{(r_1-\chi)^2+(r_2)^2}+ \frac{\mu_2 \ r_2}{(r_1+\chi)^2+(r_2)^2}\right)\theta_E\hat{\xi_2}.
\end{align}
So according to the deflection angle, the momentum changes in the lens coordinate as :
\begin{equation}\label{equ:167}
\vec{\alpha}=\frac{\Delta\vec{\acute{p}}}{|p|}=\frac{\Delta\vec{\acute{p}}}{p_z},
\end{equation}
After the scattering of photons by the lens, the momentum of photons changes to
\begin{equation}\label{equ:67}
     \acute{p}^\mu(\infty) = \acute{p^\mu}+\Delta\vec{\acute{p}}=\left(\begin{matrix}
        \gamma p_z \\
        -\beta_x \gamma p_z  \ - p_z \alpha_x\\
        -\beta_y \gamma p_z  \  -p_z \alpha_y \\
        p_z
      \end{matrix}\right),
   \end{equation}
where "$\infty$" represents the four-momentum of photons with respect to the source coordinate after the scattering of photons by the lens and $\alpha_x$ and $\alpha_y $ are the components of the deflection angle along the axis of the plane.

Finally, we perform the inverse of the Lorentz transformation (by $\beta \rightarrow - \beta$) to determine the four-momentum components of photons compared to the source coordinate. 
\begin{align}\label{equ:70}
  & p^\mu(\infty)=\notag\\
  &p_z  \left(\begin{matrix}
          \gamma & \beta_x \gamma & \beta_y\gamma & 0 \\
           \beta_x\gamma & 1+\frac{(\gamma-1)\beta_x^2}{\beta^2} & \frac{(\gamma-1)\beta_x \beta_y}{\beta^2} & 0 \\
          \beta_y\gamma & \frac{(\gamma-1)\beta_x \beta_y}{\beta^2} & 1+\frac{(\gamma-1)\beta_y^2}{\beta^2} & 0 \\
          0 & 0 & 0 & 1
        \end{matrix}\right) \left(\begin{matrix}
        \gamma \\
        -\beta_x \gamma \ - \alpha_x \\
        -\beta_y \gamma  \  -\alpha_y \\
        1
      \end{matrix}\right)
 \end{align}
which is
\begin{align}\label{equ:71}
     & p^\mu(\infty)=p_z \left(\begin{matrix}
     1-  \gamma( \alpha_{x} \beta_x +\alpha_{y} \beta_y)\\
      \frac{-\beta_x(\gamma-1)}{\beta^2}(\beta_x\alpha_x+\beta_y\alpha_y)-\alpha_x\\
        \frac{-\beta_y(\gamma-1)}{\beta^2}(\beta_x\alpha_x+\beta_y\alpha_y)-\alpha_y\\
        1
      \end{matrix}\right)
 \end {align}

Therefore, subtracting  the elements of the four-momentum from the initial value, the changes of the four-momentum obtain as

\begin{equation}\label{equ:72}
 \Delta p^\mu= p^\mu(\infty)-  \left(\begin{matrix}
     p_z \\
     0\\
     0\\
      p_z
 \end{matrix}\right)
\end{equation}
This shows that $\frac{\Delta E}{E}=- \alpha_{x}  \gamma\beta_x  -\alpha_{y}  \gamma \beta_y $ where we assume non-relativistic velocities of stars inside the galaxy, ($\gamma\simeq1$).
Therefor, the  relative frequency change of photons is
\begin{equation}\label{equ:74}
   \frac{\Delta \nu_j}{\nu_j}=-\beta_x \alpha_{x,j} -\beta_y \alpha_{y,j}
\end{equation}
where $j$ represents the label of the image from the lensing. 
In order to study the time variation of the spectral line, we can define the relative overall frequency shift in the light curve weighted by the flux of each image to measure the overall shift of the spectral barycenter, as follows:
\begin{equation}\label{equ:58}
\frac{\Delta\nu}{\nu}=\sum_j |A_j|\frac{\Delta\nu_j}{\nu_j},
\end{equation}
The total frequency shift, according to equation (\ref{equ:58}), is the frequency shift of each image, normalized to the corresponding magnification of the image as
\begin{equation}\label{equ:75}
 \frac{\Delta\nu}{\nu}=-\frac{1}{A_{total}}\sum_j |A_j|(\beta_x \alpha_{x,j} +\beta_y \alpha_{y,j}),
\end{equation}
where substituting the deflection angle for the photons, 
\begin{align}
  \frac{\Delta\nu}{\nu} =-& \frac{1}{A_{total}}\frac{ D_s}{D_{ds}} \theta_E \sum_j |A_j|  \notag \\ 
  \times&\left[\beta_x \left(\frac{\mu_1 (r_{1,j} - \chi)}{(r_{1,j} -\chi)^2 + (r_{2,j}) ^2}+\frac{\mu_2(r_{1,j}+\chi)}{(r_{1,j}+\chi)^2 + (r_{2,j})^2} \right)\right. \notag \\
  +&\left.  \beta_y \left(\frac{\mu_1 r_{2,j}}{(r_{1,j} -\chi)^2 + (r_{2,j})^2}+ \frac{\mu_2 r_{2,j}}{(r_{1,j} +\chi)^2 + (r_{2,j})^2} \right)\right].
\end{align}
Using the characteristic values for the lensing, the frequency shift is 
\begin{align}\label{equ:76}
    \frac{\Delta\nu}{\nu} &= -9.81 \times 10^{-9}\left( \frac{D_s}{8 \textrm{ kpc}}\right)^{\frac{1}{2}} \left( \frac{D_d}{4 \textrm{ kpc}}\right)^{-\frac{1}{2}} \left( \frac{D_{ds}}{4 \textrm{ kpc}}\right)^{-\frac{1}{2}} \left( \frac{M}{ M_{\odot}}\right)^{\frac{1}{2}} \notag \\ 
    \times& \sum_j \frac{|A_j|}{A_{total}}  \left[\beta_x \left(\frac{\mu_1 (r_{1,j} - \chi)}{(r_{1,j} -\chi)^2 + (r_{2,j})^2}+\frac{\mu_2(r_{1,j}+\chi)}{(r_{1,j}+\chi)^2 + (r_{2,j})^2} \right) \right. \notag \\
 + &\left. \beta_y \left(\frac{\mu_1 r_{2,j}}{(r_{1,j} -\chi)^2 + (r_{2,j})^2}+ \frac{\mu_2 r_{2,j}}{(r_{1,j} +\chi)^2 + (r_{2,j})^2} \right) \right].
\end{align}

For any position of the source star, we can identify the location of the images from the numerical solution of the lens equation. The location of images of the $j$-th image is given by $r_j = (r_{x,j},r_{y,j})$ where depending on the location of the source on the lens plane, the number of images are either $3$ or $5$. That depends on the geometrical configuration of caustic lines and the relative location of the source star. 
\begin{figure}
\begin{center}
\includegraphics[width=\columnwidth]{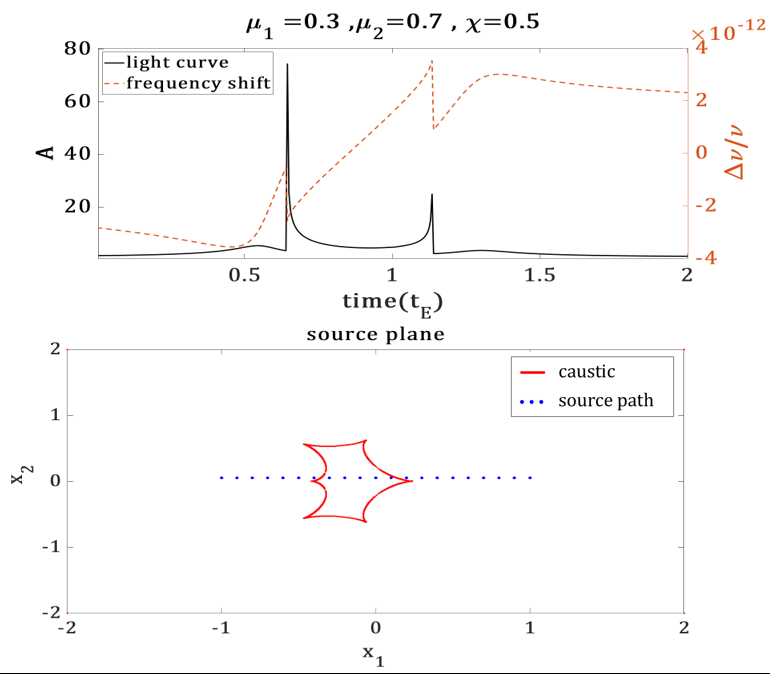}
\includegraphics[width=\columnwidth]{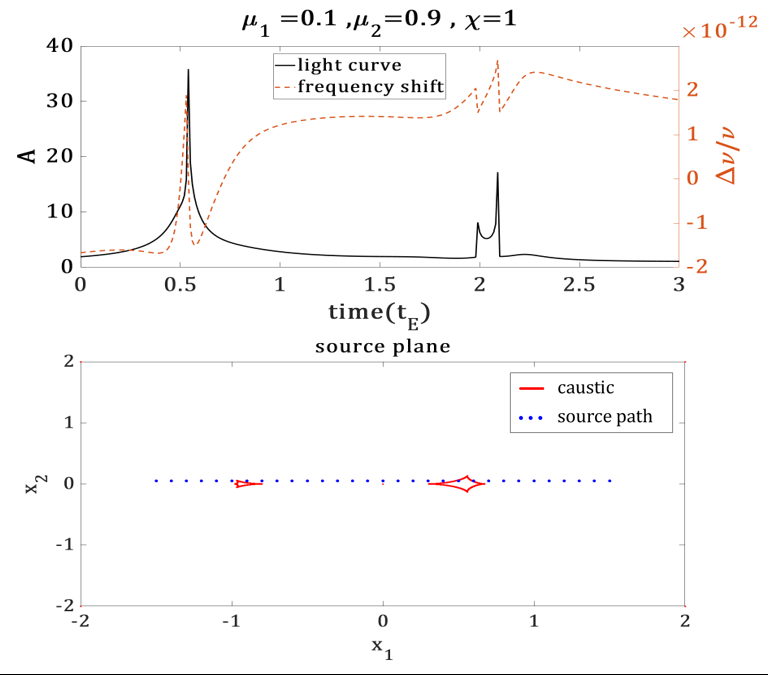}
\end{center}
\caption{ In these events, we have two lenses with mass ratios of $\mu_2 \ \textrm{ and } \mu_1 $ and their normalized distance to Einstein radius from the origin is $\chi$. The source relative velocity to the lenses is $150  km/s$. In each case , the  bottom panel represents the caustic line and relative trajectory of the source and on the top panel, the dashed line shows the frequency shift curve in terms of time and the solid line shows the light curve in terms of time.
\label{sample}}
\end{figure}

\begin{figure}
\begin{center}
\includegraphics[width=\columnwidth]{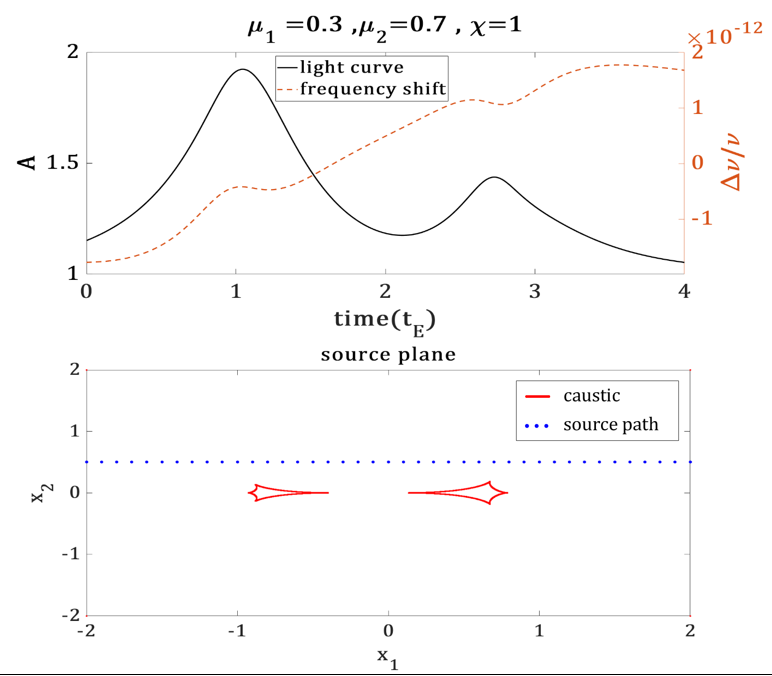}
\includegraphics[width=\columnwidth]{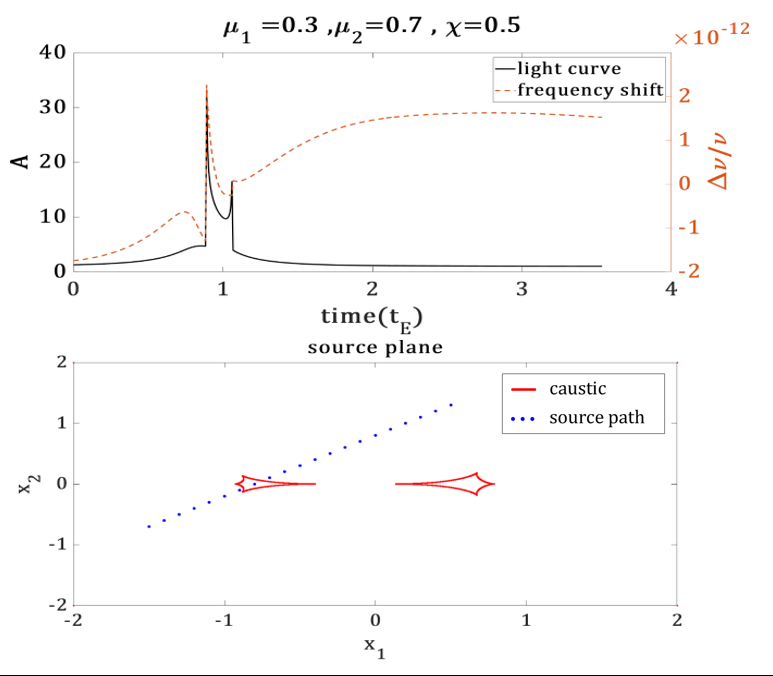}
\end{center}
\caption{ In following of Figure \ref{sample}, here we have some examples of frequency shift in which the bottom panel represents the caustic line and relative trajectory of the source and on the top panel, the dashed line shows the frequency shift curve in terms of time and the solid line shows the light curve in terms of time.
\label{sample2}}
\end{figure}
A few samples of light curves with the corresponding frequency shift are shown in Figures (\ref{sample}) and (\ref{sample2}). 
In this figures, the caustic lines and trajectory of the source on the lens plane, as well as the light curve of microlensing (black-solid curves) and frequency shift (the dashed-red curves) are shown. \\
From equation (\ref{equ:76}) frequency shift is a function of $D_s , D_d ,M ,\mu_1,\mu_2 $ and $\vec{v_{ls}} $ which ultimately adds a new constraint to Einstein time (i.e $t_E=t_E (\theta_E , u_{ls} )$) \\

For the galactic microlensing, the source is located at the galactic bulge and the lenses are located in the galactic disk. The average velocity of the objects inside the disk relative to the galaxy's center is about $200$ km/s, and the velocity of the stars in the bulge of the galaxy is about $100$ km/s. The Earth is located on one of the arms of the Milky Way galaxy and has a velocity of about $200$ km/s relative to the center of the galaxy \citep{Moniez_2017}. Considering the relative velocity of the source and the lenses, we can calculate the Einstein crossing time from equation (\ref{equ:1}). Here we adapt the source distance in $D_s=8 \textrm{kpc} $ and lens at $D_d=4\textrm{kpc}$, the total mass of lenses $ M=M_{\bigodot} $ which provides  $t_E\sim 50$ day.  One of the features of frequency shift unlike the magnification curve is that it is sensitive to the large impact parameters. This means that we can detect the ongoing microlensing events much before the magnification of the light curve happens.\\
From the Figures ( \ref{sample}) and (\ref{sample2}), this frequency shift is of the order of $10^{-12}$. For the binary lenses, due to the caustic curve, two or more peaks are seen in its light curve. In the same way, peaks or jumps happen in the frequency shift . The distance between these peaks is related to various factors, such as the distance between the two lenses and the direction of movement of the source relative to the caustic. As the distance between the two lenses decreases, the time interval between the two peaks in the light curve and in the frequency shift becomes shorter.
\section{Feasibility of observation of frequency shift in binary lensing}\label{subsec:3-5}
 
\begin{figure}
\begin{center}
\includegraphics[width=\columnwidth]{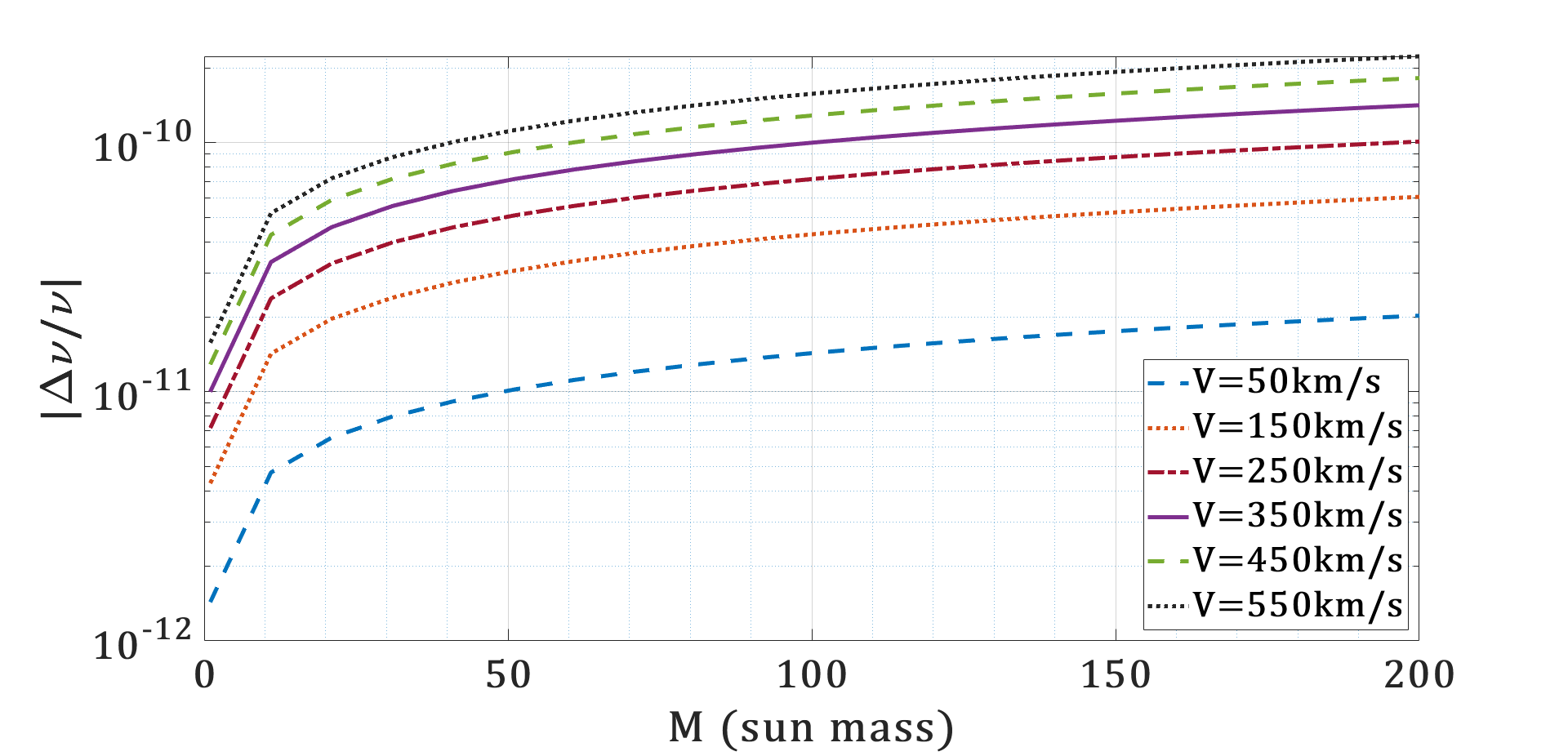}
\end{center}
\caption{Frequency shift in term of total mass of lenses for various relative transverse velocities.
\label{fig:501}}
\end{figure}

\begin{figure}
\begin{center}
\includegraphics[width=\columnwidth]{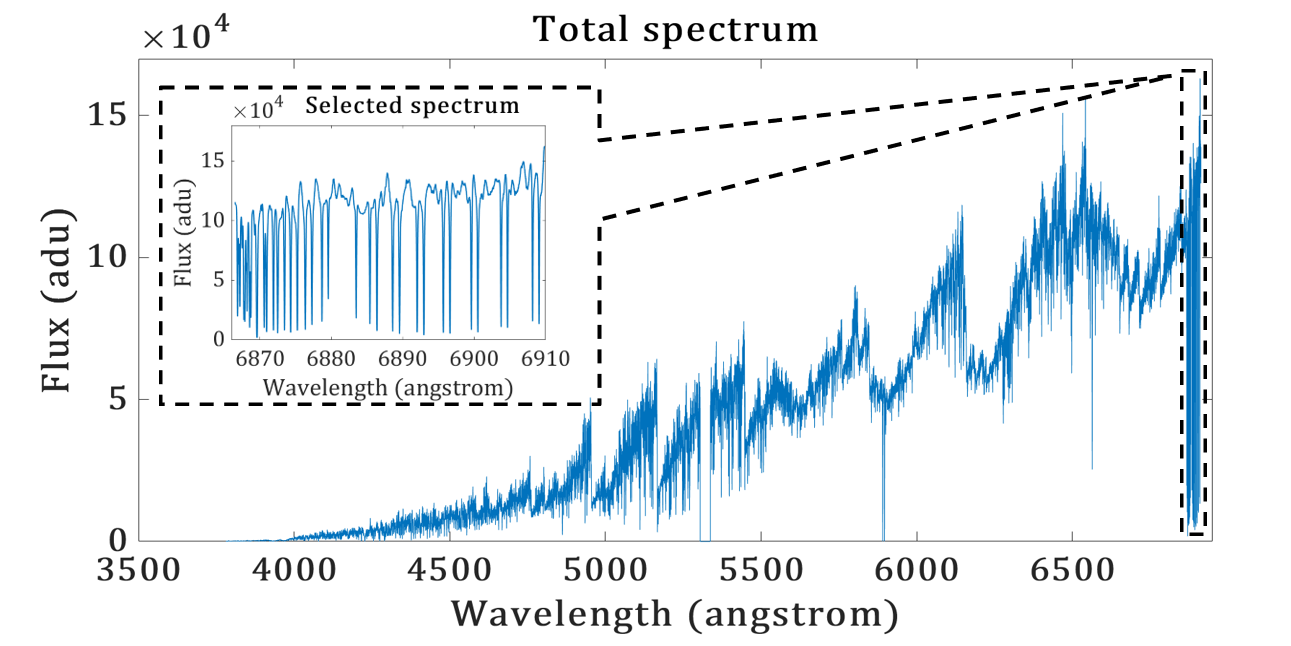}
\end{center}
\caption{Spectra of Alpha Orionis and the absorption lines in the part of the spectrum selected to study the frequency shift. Data is adapted from ESO archive \label{fig:18}}
\end{figure}

\begin{figure}
\begin{center}
\includegraphics[width=\columnwidth]{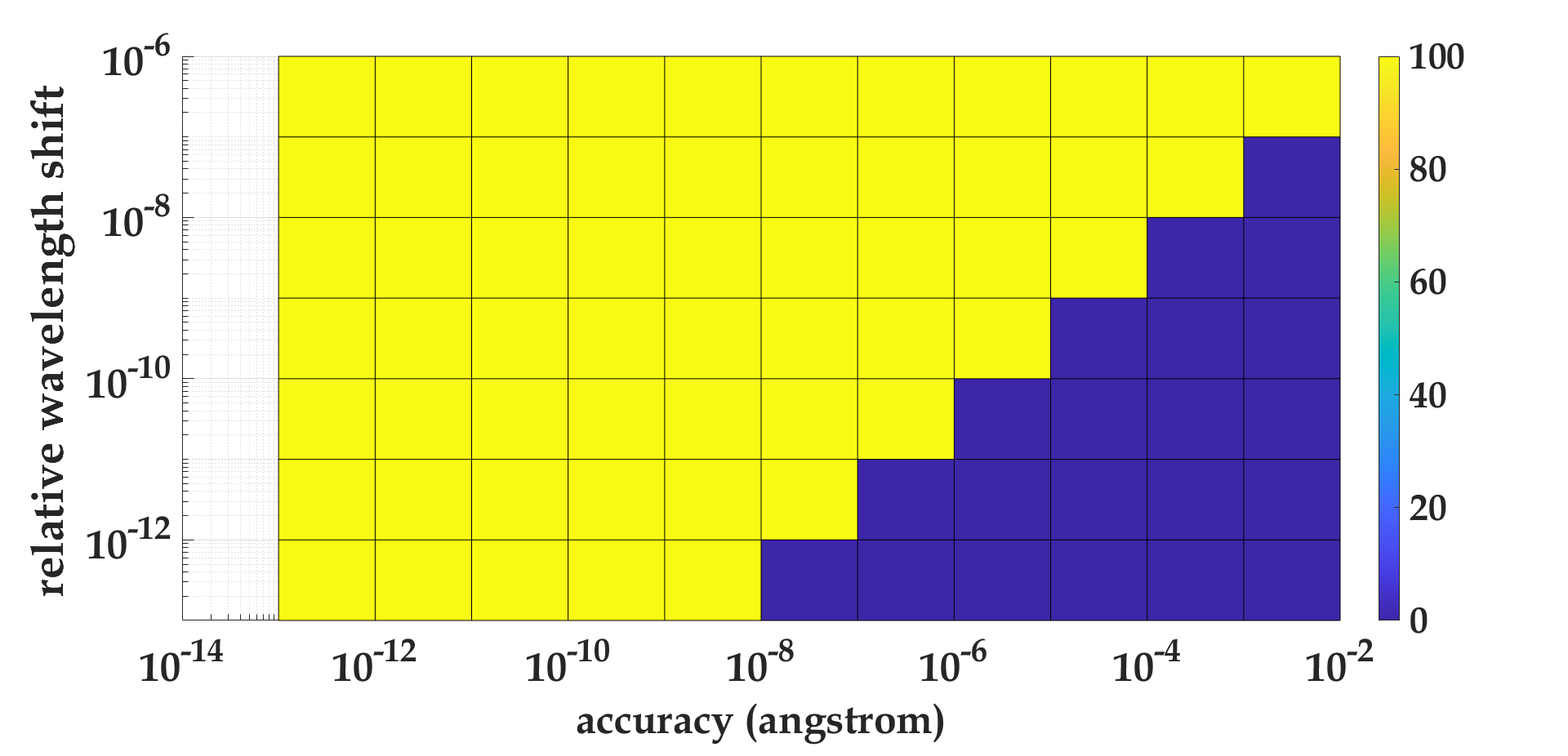}
\end{center}
\caption{ The x-axis represents different wavelength accuracies, and the y-axis shows the relative wavelength shift .In this figure, color-coded regions correspond to the percentage of acceptable shift.  
\label{fig:19} }
\end{figure}

In this section we discuss on feasibility of frequency shift measurement . So far, we have seen that the frequency shift ranges from $10^{-10}$ to $10^{-12}$ for stellar mass objects. Figure (\ref{fig:501}) represents the frequency shift in terms of the total mass of two lenses for various relative transverse velocities.
In order to study the feasibility of the frequency shift, we perform a simulation with two spectra from single source star: the first one is the original spectra of the star and the second one is the the shifted spectra as a result of frequency shift during the microlensing. 
We choose the spectrum of Betelgeuse (i.e Alpha Orionis ) from the ESO archive \footnote{Based on observations collected at the European Southern Observatory under ESO program 60.A-9501 and data obtained from the ESO Science Archive Facility with DOI(s) under https://doi.org/10.18727/archive/33} .(see Figure (\ref{fig:18})). Then we simulate the second spectrum according to the frequency shift for the binary lenses. In order to have realistic realisation of the data, for each data point in the simulated spectrum we take into account the error bar of the spectrometer and impose extra shift with a normal distribution.
As the first attempt, we subtract the two spectrum in terms of wavelength. In another word let take $\lambda_i$ as the $i$th data point and $\lambda'_i$ from the second spectrum, we define $\delta\lambda_i = |\lambda_i - \lambda'_i|$. Then calculate the average value of $\delta\lambda_i$ and compare it with the initial frequency shift in the simulation. This gives us a criteria to test measurability of a frequency shift with a spectrometer. Having accurate spectrometer by means of accuracy in measuring the wavelength,  improves the frequency shift measurement. Figure (\ref{fig:19}) represents the relative frequency shift for a range of spectrometer accuracy within the range of $10^{-13}$\AA~  to $10^{-2}$\AA. For bin in this figure we perform $100$ realisations from the simulation and compare the result with the initial value. The value of each bin represents the percentage of cases that frequency shift is consistent with the initial value in the simulation.   
From the Figure (\ref{fig:19}), the colored area is the region that we can measure the frequency shift. Our desire in this observation is measuring the frequency shift within the range of $[10^{-12},10^{-10}]$ which requires the accuracy within the range of almost  $[10^{-7},10^{-5}]$. 

\begin{figure}
\begin{center}
\includegraphics[width=\columnwidth]{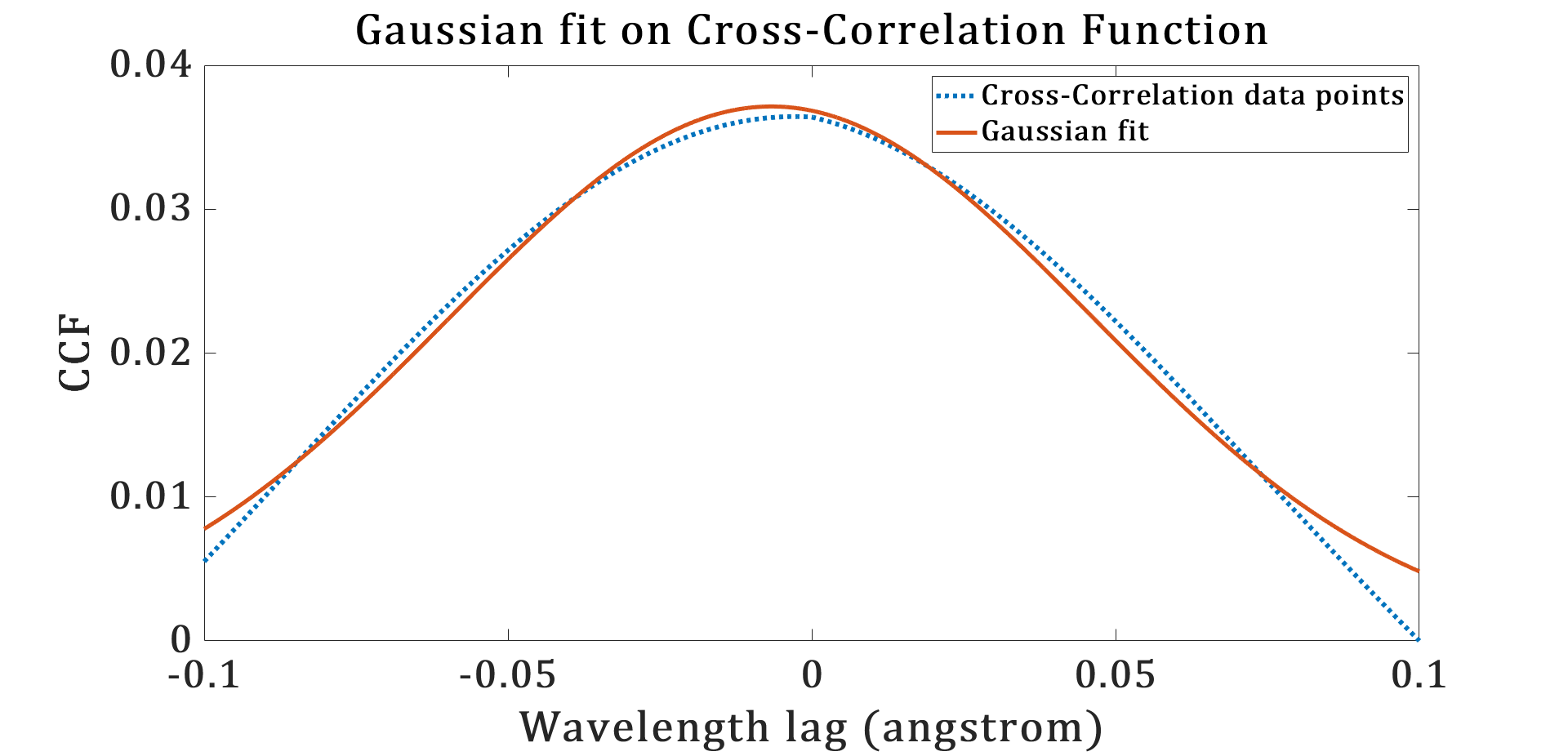}
\end{center}
\caption{The cross-corrolation in y-axis as a function of distance between two spectra in terms of wavelength in x-axis. We simulate a relative shift of  $\delta\lambda/\lambda\sim -10^{-6}$. The blue line represents the cross-correlation  value and the red curve is the best Gaussian fit to the blue curve.  The best value for the peak of this function is $\delta\lambda=- 0.00688 \pm 8.1898\times 10^{-6}$\AA  ~ which provides a relative wavelength shift of $\delta\lambda/\lambda =-9.6312 \times 10^{-7}  \pm 1.1893 \times 10^{-9}$
\label{fig:522}
}
\end{figure}

We also use the the second method for comparing the two spectrum with the cross-correlating data points.  
The cross-correlation techniques for measuring velocity shifts is well-established, with seminal work by \citep{SIMKIN1972} and \citep{1979AJ.....84.1511T} . 
The advancements in cross-correlation also can be found in  \citep{1994ApJ...420..806Z,1995AJ....109.1371S,2007ApJ...662..602T,10.1046/j.1365-8711.2003.06633.x} 
The cross correlation is the most popular procedure for deriving relative velocities between the two spectrum. This technique uses of all the available information in the two spectra, and has proven to be far superior to simply comparing the Doppler shifts between the central wavelengths of lines when the S/N is low \citep{Allende_Prieto_2007} 
The cross-correlation function using two discrete quantities defines as follows 
\begin{equation}
\label{equ:81}
 f(l)=\sum_l S_i S'_{i+l}
\end{equation}
In which $S$ is spectrum data points and $ {S'}$ is shifted spectrum. We multiply each $i$-th component of the first spectrum by the $( i + l)$th component of the shifted spectrum and $f(l)$ according to equation (\ref{equ:81}) obtained by summing over all the terms. The maximum value of  $f(l)$  results in the  amount of shift. According to \cite{Allende_Prieto_2007} we fitted a Gaussian function on $f(l)$ and by estimating the peak of this Gaussian function, we derive the frequency shift value (see Figure \ref{fig:522}).

We note that since our data points are discrete, instead of using these data, we perform pre-analysis procedure by means of interpolating between two consecutive data points. Then we use these new data for cross-correlation function.  While this method uses all the possible information of the spectrum, we faced with the lack of computational power so we could use interpolating data points with up to the cadence of  $ \sim 10^{-6} $\AA .

\begin{figure}
\begin{center}
\includegraphics[width=\columnwidth]{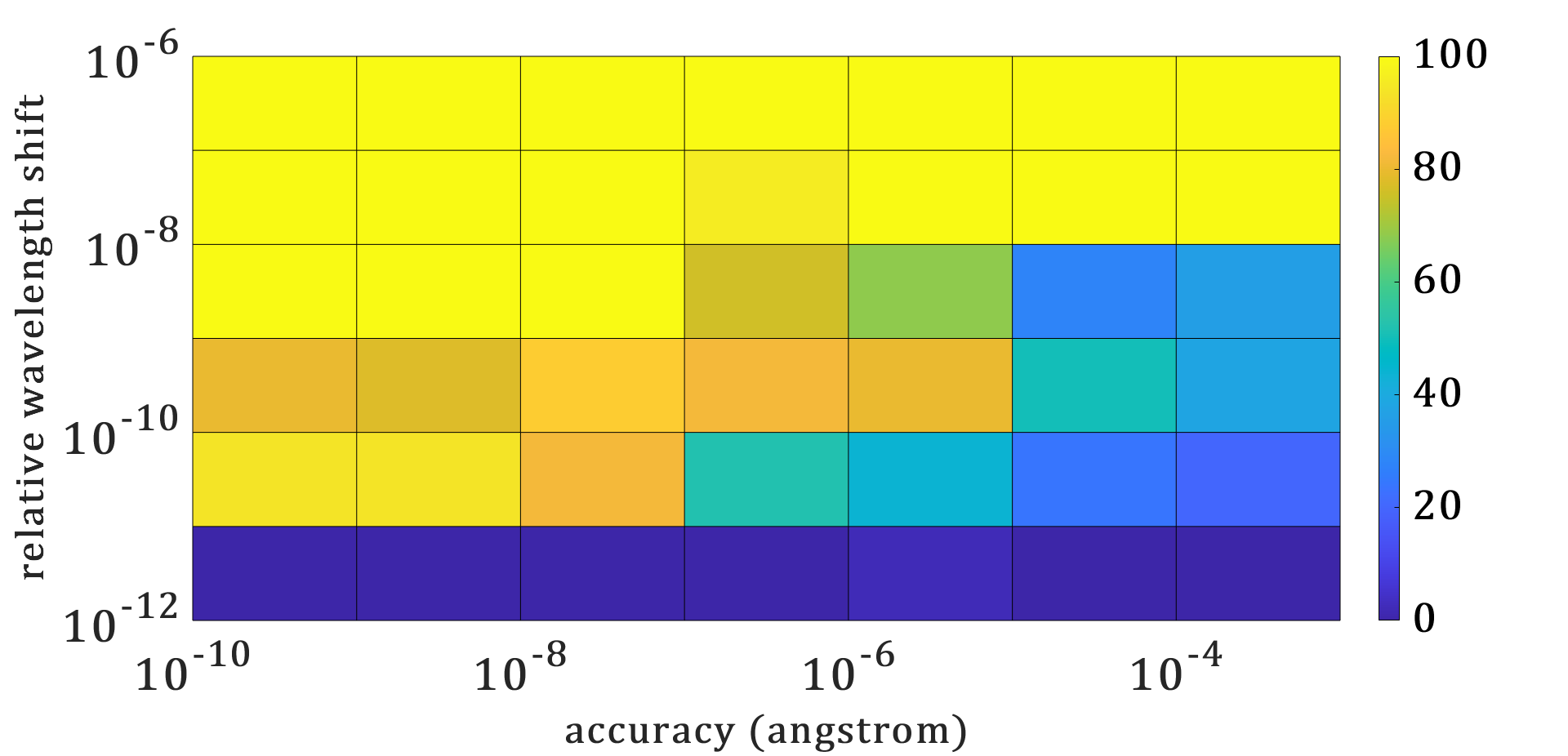}
\end{center}
\caption{ The percentage of success in obtaining the frequency shift by the cross correlation method \label{fig:20}}
\end{figure}

The rest of our analysis is similar to the first method where for a given spectroscopic resolution and the frequency shift we generate $100$ realisation of the shifted spectra and measure the frequency shift with the cross-correlation function. Figure \ref{fig:20} shows the percentage of successful recognition of the frequency shift. Having a powerful computational machine, we can push the lower-band of the frequency shift to the smaller values. Comparing Figure (\ref{fig:19}) with  (\ref{fig:20} )
shows that cross correlation method can identify the frequency shift with higher probability.

\section{Conclusion}\label{4}
\label{subsec:4-1}
In this paper, we extended the formalism of the frequency shift for the single lens \citep{Rahvar_2020} to the binary lensing. The observational feature due to the transverse velocity of the lens is a frequency shift in the spectra of the source star in the gravitational microlensing. This extra information can partially break the degeneracy between the lens parameters in single as well as binary lensing. We also derived the frequency shift from the Shapiro time delay formalism which is consistent with \citep{Rahvar_2020} from the Lorentz boost formalism. For the binary lenses, we have shown that frequency shift  is in the same order as the single lens. In fact, frequency shift for binary lensing varies from $\sim10^{-12}$ to $\sim10^{-10}$ depending on the mass of the lenses and the relative transverse velocity of the lenses to the source.

In order to test the feasibility of the frequency shift observations in the microlensing, we simulated the affect of frequency shift on the spectrum of source star taking into account the spectral resolution of the instrument . Then we compare the averaged value of frequency shift with the initial value . We also introduced cross-correlation method for investigating frequency shifts, by examining all data points in the spectrum. This approach can reasonably estimate the spectral shift.

For the observation of frequency shift our plan is to perform follow-up observation of ongoing microlensing events with spectroscopic observations .This will open a new window for gravitational microlensing observations especially for exoplanet studies to have extra constrain on the parameters of lens and planets orbiting around the lens star .


\section*{Data Availability}

No data is generated in this work.



\bibliographystyle{mnras}



\appendix
\section{ frequency shift for single lens by special relativity } \label{app:1}
\subsection{Lensing equation for Microlensing by Single lens}\label{subsec:2-1}
\begin{figure}
\begin{center}
\includegraphics[width=\columnwidth]{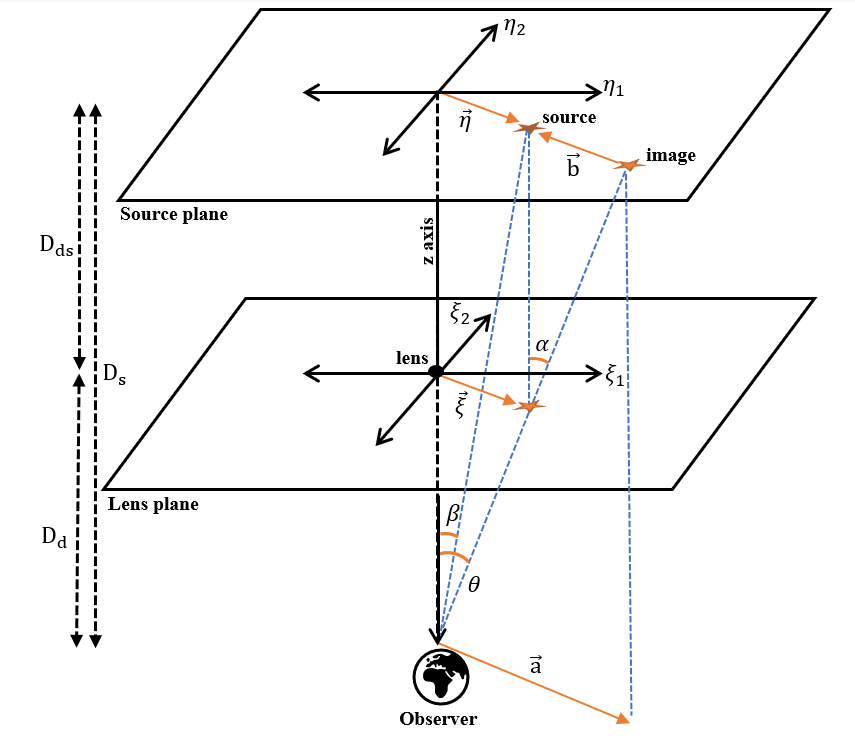}
\end{center}
\caption{Overview of the location of the observer and a single lens in the observer, lens, and source coordinate systems.
\label{fig:2}}
\end{figure}

The configuration of lensing with source, lens and observer with the corresponding planes perpendicular to the light ray
is presented in Figure (\ref{fig:2}).
Here, $ D_s $ is the distance of the source from the observer, $ D_d $  is the distance of the lens from the observer and  $ D_{ds} $ is the distance of the source from the lens. According to the geometry of the light deflection in Figure (\ref{fig:2}), we write the following equation for the vectors $ \Vec{\eta }$,  $ \Vec{b }$ and $ \Vec{a }$ :
\begin{equation}\label{equ:7}
   \vec{b}= \vec{\eta}-\vec{a}
\end{equation}
Where $\vec{\eta}$ is the source position on the source plane, $\vec{b}$ is the vector from the image to the source position on the source plane and $ \vec{a}$  is the position of the image projected on the source plane. We can write this equation in terms of the small angles and distances as in Figure (\ref{fig:2}),
\begin{equation}\label{equ:8}
   D_{ds}\vec{\alpha}=D_s \vec{\beta}-D_s\vec{\theta},
\end{equation}
where  $ \vec{\beta}$ is the angular position of the source and $ \vec{\theta}$ is the angular position of images, $ \vec{\alpha}$ is the deflection angle and according to the light bending in Schwarzschild metric. The deflection angle for multiple lenses is given by:
\begin{equation}\label{equ:9}
   \vec{\alpha}=\sum_i \frac{4GM_i}{c^2D_d}\frac{\vec{\theta}-\vec{\theta_i}}{|\vec{\theta}-\vec{\theta_i}|^2}
\end{equation}
Index $i$ refers to the point-like lenses and $ \vec{\theta_i }$ is angular position of $i$-th lens. In single-lens mode, we have one lens at the origin of the coordinate systems and the deflection angle is given by:
\begin{equation}\label{equ:10}
   \vec{\alpha}=\frac{4GM}{c^2D_d}\frac{\vec{\theta}}{|\vec{\theta}|^2}.
\end{equation}
The Einstein angle which is the angular size of the Einstein ring\footnote{The Einstein ring occurs only when the observer, the lens, and the source located along a line} is defined as
\begin{equation}\label{equ:11}
   \theta_E^2 =\frac{4GMD_{ds}}{c^2D_dD_s}
\end{equation}
substituting equation (\ref{equ:10}) in equation (\ref{equ:8}) and using the definition of Einstein angle, we obtain the lens equation as
\begin{equation}\label{equ:12}
   \theta^2 = \theta_E^2 +\beta\theta,
\end{equation}
 where this equation for the single lens has the two solutions of

 \begin{equation}\label{equ:13}
\theta^{\pm}=\frac{\beta\pm\sqrt{\beta^2+4\theta_E^2}}{2}.
\end{equation}
 The magnification factor is given as the area of images to the area of source assuming the surface brightness is constant for the image and source on the lens plane,
\begin{equation}\label{equ:14}
   A=\left|\frac{{ds}_{image}}{{ds}_{source}}\right| = \ \left|\frac{\theta \ d\theta }{\beta \ d\beta}\right|.
\end{equation}

According to equation (\ref{equ:13}) two images form by single microlensing and the total magnification during the gravitational microlensing is obtained from the sum of the two magnifications:
\begin{equation}\label{equ:16}
   A=|A_1|+|A_2|.
\end{equation}
where the total magnification is
\begin{align}\label{equ:17}
   A=\left|\frac{1}{2} \ \left(1+ \frac{\beta^2+2\theta_E^2}{\beta\sqrt{\beta^2+4\theta_E^2}}\right)\right|+&\left|\frac{1}{2} \ \left(1- \frac{\beta^2+2\theta_E^2}{\beta\sqrt{\beta^2+4\theta_E^2}}\right)\right|
\end{align} 
By normalizing $ \beta $ with Einstein angle (i.e . $ \bar{\beta}=\frac{\beta}{\theta_E} $ ), equation (\ref{equ:17}) simplifies to
\begin{equation}\label{equ:18}
   A=\frac{\bar{\beta}^2+2}{\bar{\beta}\sqrt{\bar{\beta}^2+4}}.
\end{equation}
\subsection{Frequency shift in Microlensing by single lens}
The concept of the gravity-assist is based on the momentum transfer to a particle while scattering from a planet orbiting around the Sun. While the magnitude of the incoming and outgoing momentum of the scattered particle with respect to the planet doesn't change, the overall momentum of the particle with respect to the Sun, depending on the relative motion of the particle and planet can change. This method was proposed by Yuri Kondratyuk in 1938  for sending spacecraft to the other planets of the solar system \citep{harvey2007russian} and even changing the orbit of celestial objects \citep{Rahvar2023}. Now, we apply the gravity assist to the photons. 
\begin{figure}
\begin{center}
\includegraphics[width=\columnwidth]{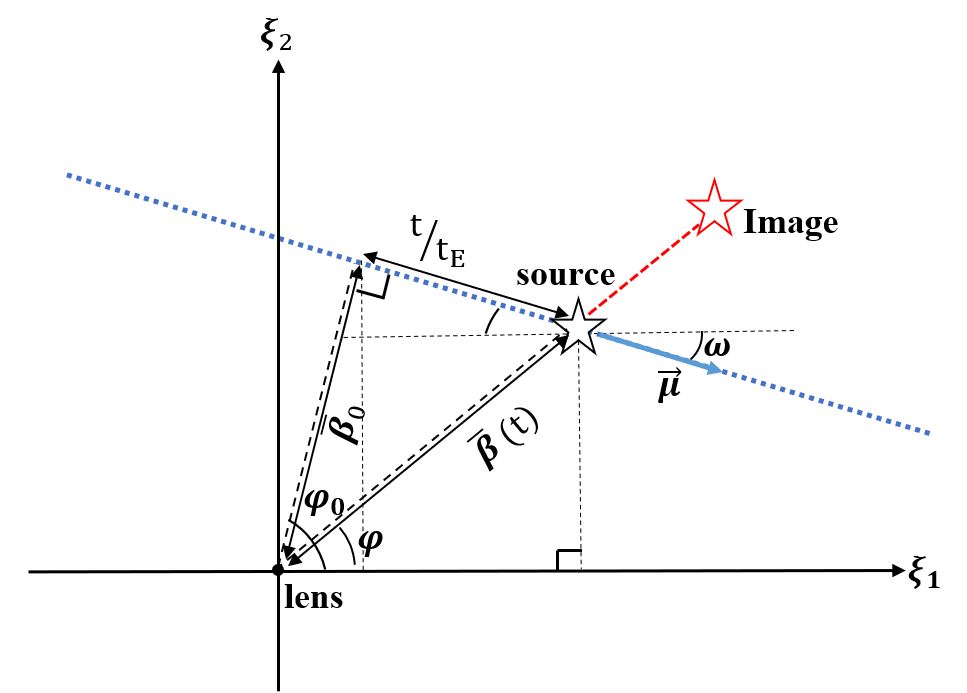}
\end{center}
\caption{ The relative transverse motion of the source with respect to the lens on the lens plane. The lens is located at the center of the coordinate system and the observer is along the z-axis. The angle of $\omega$ is the relative angle between the velocity of the source and $\xi_1$ axis. $\phi$ is defining as the angle of source position and $\xi_1$ axis. \label{fig:9}}
\end{figure}
According to Figure (\ref{fig:2}), let us assume photons are emitted from the source along the z-direction. Therefore, the four-momentum of the photon in the source's coordinate system can be written as follows:
\begin{equation}\label{equ:41}
p^\mu =\left(\begin{matrix}
        p_z \\
        0 \\
        0 \\
        p_z
      \end{matrix}\right),
\end{equation}
where the energy of the photon is $p_0 = p_z=\frac{E}{c}$.
We take the horizontal axis of coordinate systems parallel to the transverse velocity of the source. We define the relative velocity of the source with respect to the lens by $V_{sl}$, then $V_{ls}=-V_{sl}$. From figure \ref{fig:9} $\vec{V}_{sl} = \vec{V}_{sl}\cos\Omega \ \hat{\eta_1}+ \vec{V}_{sl}\sin\Omega \ \hat{\eta_2}$
\\
The Lorentz transformation matrix between the lens and source frames is 
\begin{equation}\label{equ:42}
\Lambda = \left(\begin{matrix}
          \gamma & -\beta_x \gamma & -\beta_y\gamma & 0 \\
           -\beta_x\gamma & 1+\frac{(\gamma-1)\beta_x^2}{\beta^2} & \frac{(\gamma-1)\beta_x \beta_y}{\beta^2} & 0 \\
          -\beta_y\gamma & \frac{(\gamma-1)\beta_x \beta_y}{\beta^2} & 1+\frac{(\gamma-1)\beta_y^2}{\beta^2} & 0 \\
          0 & 0 & 0 & 1
        \end{matrix}\right).
\end{equation}
where $\beta=\frac{\vec{V}_{ls}}{c} \quad
 \beta_x =\beta\cos\Omega \ , \quad \beta_y = \beta\sin\Omega    $ 
and after applying the Lorentz transfer matrix, the four-momentum of the photon in the lens coordinate system is as follows:
\begin{equation}\label{equ:43}
\acute{p^\mu}=\left(\begin{matrix}
        \gamma p_z\\
        -\beta_x \gamma p_z \\
        -\beta_y \gamma p_z \\
        p_z
      \end{matrix}\right).
\end{equation}
Using the deflection angle 
$\vec{\alpha}=-\frac{4GM}{c^2D_d}\frac{\vec{\theta}}{|\vec{\theta}|^2},
$
and substituting the position images, according to  
Figure (\ref{fig:9}), the deflection angle is given by 
\begin{equation}\label{equ:45}
\vec{\alpha}=-|\vec{\alpha}|( \cos\varphi \ \hat{\xi_1}+\sin\varphi \ \hat{\xi_2}),
\end{equation}

where the negative sign in the above equation means that the light is deflected towards the lenses. So according to the deflection angle, the momentum changes in the lens coordinate is as follows:
\begin{equation}\label{equ:46}
\vec{\alpha}=\frac{\Delta\vec{\acute{p}}}{|p|}=\frac{\Delta\vec{\acute{p}}}{p_z},
\end{equation}
In other words, 
\begin{equation}\label{equ:47}
\Delta\vec{\acute{p}}=-p_z|\alpha|( \cos\varphi \ \hat{\xi_1}+\sin\varphi \ \hat{\xi_2}).
\end{equation}

Therefore, the four-momentum in the lens coordinate system after applying gravitational deflection by a single lens is given as:
\begin{equation}\label{equ:48}
\acute{p}^\mu(\infty) = \acute{p^\mu}+\Delta\vec{\acute{p}}=\left(\begin{matrix}
        \gamma p_z \\
        -\beta_x \gamma p_z  \ - p_z |\alpha|\cos\varphi\\
        -\beta_y \gamma p_z  \  -p_z |\alpha|\sin\varphi \\
        p_z
      \end{matrix}\right),
\end{equation}
After using the Lorentz transformation with $-\beta$, we return back to the source coordinate system
\begin{equation}\label{equ:49}
   p^\nu(\infty)= \Lambda_{\mu}^{\nu} \acute{p}^\mu(\infty)
 \end{equation}
If we consider $\alpha_x=|\alpha|\cos\varphi$ and $\alpha_y=|\alpha|\sin\varphi$ , equation \ref{equ:49} will have these components:
\begin{align}\label{equ:50}
    & p^\mu(\infty)=p_z \left(\begin{matrix}
      1- \gamma (  \alpha_{x} \beta_x +\alpha_{y} \beta_y)\\
      \frac{-\beta_x(\gamma-1)}{\beta^2}(\beta_x\alpha_x+\beta_y\alpha_y)-\alpha_x\\
        \frac{-\beta_y(\gamma-1)}{\beta^2}(\beta_x\alpha_x+\beta_y\alpha_y)-\alpha_y\\
        1
      \end{matrix}\right)
 \end{align}
 Before comparing this with initial four-momentum , it is important to check it from the perspective of an observer on Earth .
 Considering that the plane of the observer is parallel to the plane of the source , we assume that the velocity of the Earth with respect to the source may have an angle $\Theta $. Therefore, for the simplicity ,we rotate the coordinates of the source by the amount of $\Theta$ . We do this in order to use only the Lorentz transformations for the velocity in one direction. Therefore, the four-momentum in source coordinate will be as follows:
 \begin{align}\label{equ:250}
    & p^\mu= \left(\begin{matrix}
         p_0\\
         p_x\cos{\Theta}\\
        p_y\sin{\Theta}\\
        p_z
      \end{matrix}\right)
 \end{align}
 With the Lorentz transformations, we get the four-momentum from the point of view of the observer on the Earth moving with speed of $V_{earth}$ and $\Tilde{\beta}=\frac{V_{earth}}{c}$
 \begin{align}\label{equ:251}
    & \Tilde{p}^\mu=\left(\begin{matrix}
         \Tilde{\gamma} & -\Tilde{\beta}\Tilde{\gamma} & 0 & 0 \\
           -\Tilde{\beta} \Tilde{\gamma}& \Tilde{\gamma}& 0 & 0 \\
          0 & 0 & 1 & 0 \\
          0 & 0 & 0 & 1
        \end{matrix}\right)\left(\begin{matrix}
         p_0\\
         p_x\cos{\Theta}\\
        p_y\sin{\Theta}\\
        p_z
      \end{matrix}\right)
 \end{align}
So we will see that :
 \begin{equation}\label{equ:252}
    \Tilde{p}_0=\Tilde{\gamma} p_0 -\Tilde{\gamma} \Tilde{\beta} p_x \cos {\Theta}
 \end{equation}
 With this definition, the zeroth component of the four-momentum from the point of view of the Earth's observer before and after the deflection will be as follows:
  \begin{align}\label{equ:253}
&\Tilde{p}_0=\Tilde{\gamma}p_z \notag  \\ &\Tilde{p}_0(\infty)=\Tilde{\gamma} (p_z - \alpha_{x}p_z \gamma\beta_x -\alpha_{y}p_z  \gamma \beta_y )-\Tilde{\gamma} \Tilde{\beta} p_x (\infty)\cos {\Theta}
 \end{align}
 Assuming non-relativistic speeds, we could consider $\gamma=\Tilde{\gamma}\sim 1$
 So energy changes from the point of view of an observer on Earth is :
 \begin{equation}\label{equ:254}
     \frac{\Delta E}{E}=-[(\beta_x+\Tilde{\beta}\cos{\Theta})\alpha_x+\beta_y\alpha_y]
 \end{equation}
 So it seems that the movement of the earth relative to the source in the non-relativistic limit adds only one additional term to the velocity, which can be considered as a correction in the real data and obtain the frequency shift in the coordinates of the source.
 Comparing the four-momentum of photons in the initial state from equation (\ref{equ:41}) with the four-momentum in the final state due to the gravitational lensing effect and gravity assist effect, we can calculate the modification on the momentum of photons as a result of the transverse velocity of the lens. 
\begin{equation}\label{equ:51}
\frac{\Delta \nu_j}{\nu_j}=-\beta_x |\alpha|\cos\varphi_j -\beta_y|\alpha|\sin\varphi_j
\end{equation}
\begin{align}\label{equ:52}
\frac{\Delta \nu_j}{\nu_j} =- &|V_{ls}||\alpha| (\cos{\omega} \cos\varphi_j +\sin{\omega}\sin\varphi_j)=\notag \\
& |V_{ls}||\alpha| \cos{(\omega- \varphi_j)}\notag \\
\end{align}
Where index j refers to the j-th image. According to Figure (\ref{fig:9}) $\omega=\frac{\pi}{2}-\varphi_0$ and  $\varphi$ depends on time as

\begin{align}\label{equ:54}
\cos\varphi = \frac{\bar{\beta}_0\cos(\varphi_0)+\frac{t}{t_E}\sin(\varphi_0)}{\bar{\beta}(t)} \notag \\
\sin\varphi = \frac{\bar{\beta}_0\sin(\varphi_0)-\frac{t}{t_E}\cos(\varphi_0)}{\bar{\beta}(t)} 
\end{align}
Considering $\omega=\frac{\pi}{2}-\varphi_0$ , we calculate the following value
\begin{align}\label{equ:53}
\cos(\omega-\varphi )&= \frac{\bar{\beta}_0\cos(\varphi_0)+\frac{t}{t_E}\sin(\varphi_0)}{\bar{\beta}(t)}\sin{\varphi_0}\notag \\
&+\frac{\bar{\beta}_0\sin(\varphi_0)-\frac{t}{t_E}\cos(\varphi_0)}{\bar{\beta}(t)}\cos{\varphi_0}\notag\\
&=\frac{\bar{\beta}_0 \sin{2\varphi_0}+\frac{t}{t_E}\cos{2\varphi_0}}{\bar{\beta}(t)}
\end{align}

where $\bar{\beta}_0$ is the minimum impact parameter
and  the impact parameter from the geometry of the trajectory of the source depends on time as
\begin{equation}\label{equ:55}
\bar{\beta}^2(t) = \bar{\beta}_0^2+\left(\frac{t}{t_E}\right)^2
\end{equation}
where $t_E ={\theta _E}/{\mu}$ and $\bar{\beta} ={\beta}/{\theta_ E}$.

Using Equations (\ref{equ:55}) and equation (\ref{equ:54}) we can obtain the frequency shift of images as follows
 \citep{Rahvar_2020}
\begin{equation}\label{equ:56}
\frac{\Delta\nu_i}{\nu_i}=-\frac{C(t)}{|\bar{\theta}_i(t)|},
\end{equation}
where $\bar{\theta}_i(t)$ is the angular position of the i-th image normalized by the Einstein angle and
\begin{align}
 C(t)=&2.23\times 10^{-12}\left(\frac{M}{0.5 M_\odot}\right)^{\frac{1}{2}} \left(\frac{D_s}{8.5 kpc}\right)^{-\frac{1}{2}}\left(\frac{v_{ls}}{200 \frac{km}{s}}\right)\notag\\
 &\times  \left(x(1-x)\right)^{-\frac{1}{2}} \left( \frac{\bar{\beta}_0 \sin{2\varphi_0}+\frac{t}{t_E}\cos{2\varphi_0}}{\bar{\beta}(t)} \right) 
\end{align}\label{equ:57}

 For the single lens, we have two images . The positive and negative parity images from the lensing
equation result in the opposite frequency shifts where $C(t)$
for both images are the same and $\theta(t)$ is different for the
two images. We measure the total amount of frequency changes resulting from both of these images where each image is weighted by the magnification factor. 

\begin{figure}
\begin{center}
\includegraphics[width=\columnwidth]{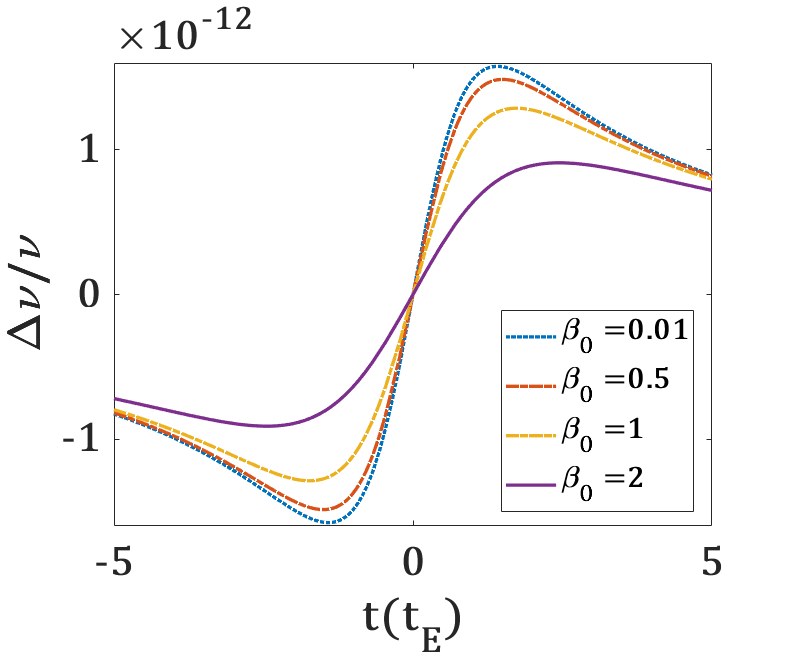}
\end{center}
\caption{ The frequency shift as a function of time (normalized to $t_E$) for a lens with the typical values adapted from equation (\ref{equ:59}). The frequency shift is plotted for four minimum impact parameters.These plots are for different values of impact parameter considering $x=0.5$ and $\varphi_0=\pi$. According to the previous assumptions, here we also imagined a single lens, which is almost halfway between us and the source with $ D_s=8.5 kpc $  \label{fig:10}}
\end{figure}

 So the relative overall frequency shift , will be as follows:
\begin{equation}
\frac{\Delta\nu}{\nu}=\sum_i |A_i|\frac{\Delta\nu_i}{\nu_i},
\end{equation}
So we will have
\begin{equation}\label{equ:59}
\frac{\Delta\nu}{\nu}=\frac{C(t)}{A}\left(\frac{|A_1|}{\bar{\theta}_1}+\frac{|A_2|}{\bar{\theta_2}}\right).
\end{equation}
Figure (\ref{fig:10}) shows the frequency change of a spectral line in the gravitational microlensing event as a function of time (normalized time by Einstein crossing time) for different values of the $\beta_0$

\section{Shapiro effect for binary lenses}\label{app:2}

\begin{figure}
\begin{center}
\includegraphics[width=\columnwidth]{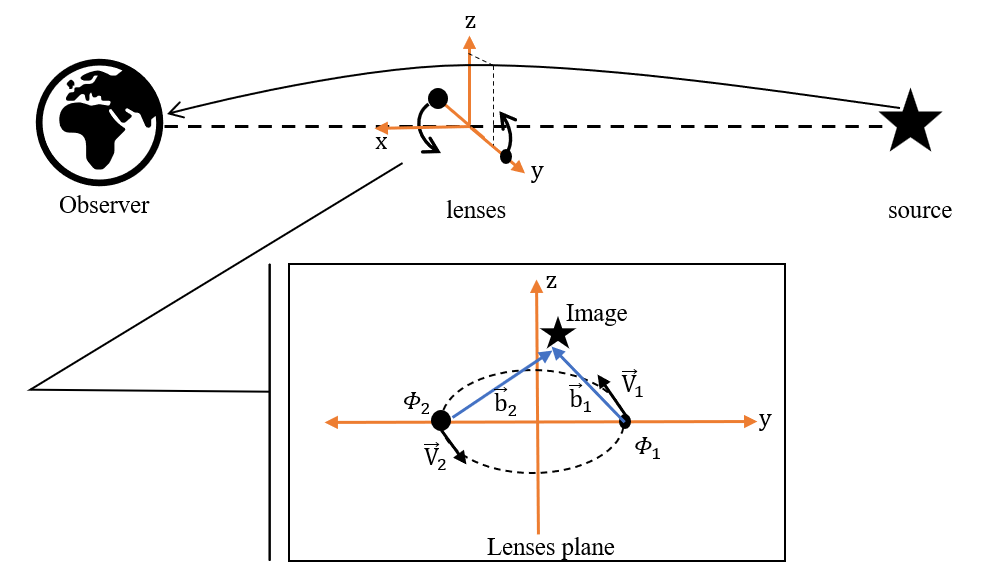}
\end{center}
\caption{ Two lenses in the binary lensing system as two gravitational potentials, with the impact of parameters $b_1$ and $b_2$  \label{fig:25}}
\end{figure}
In the case of double lenses with different speeds, we write the Shapiro delay integral again. Consider two potentials $\phi_1$ and $\phi_2$ representing two lenses  (see figure \ref{fig:25}). So frequency shift will be as following :
\begin{equation}\label{equ:100}
\frac{\delta\nu}{\nu} = 2\int (\frac{d\phi_1}{dt}+\frac{d\phi_2}{dt})dl
\end{equation}
This integral can be separated and each one can be calculated separately
\begin{equation}\label{equ:101}
\frac{\delta\nu}{\nu} = 2\int (\Vec{V_1}.\nabla\Phi_1)dl+2\int (\Vec{V_2}.\nabla\Phi_2)dl
\end{equation}
so we will have :
\begin{equation}\label{equ:102}
\frac{\delta\nu}{\nu} = -\frac{4GM}{b_1} \hat{b_1}.\Vec{V_1}-\frac{4GM}{b_2} \hat{b_2}.\Vec{V_2}
\end{equation} 
To calculate the frequency shift related to binary lenses, first we should find the position of the images and then for each of them, find the vector $\Vec{b_i}$ with respect to the lenses according to Figure 1. In this way, the total frequency shift value can be obtained from the sum of the frequency shift of the images according to their magnification.

\section{Orbital velocity }\label{app:3}
\begin{figure}
\begin{center}
\includegraphics[width=\columnwidth]{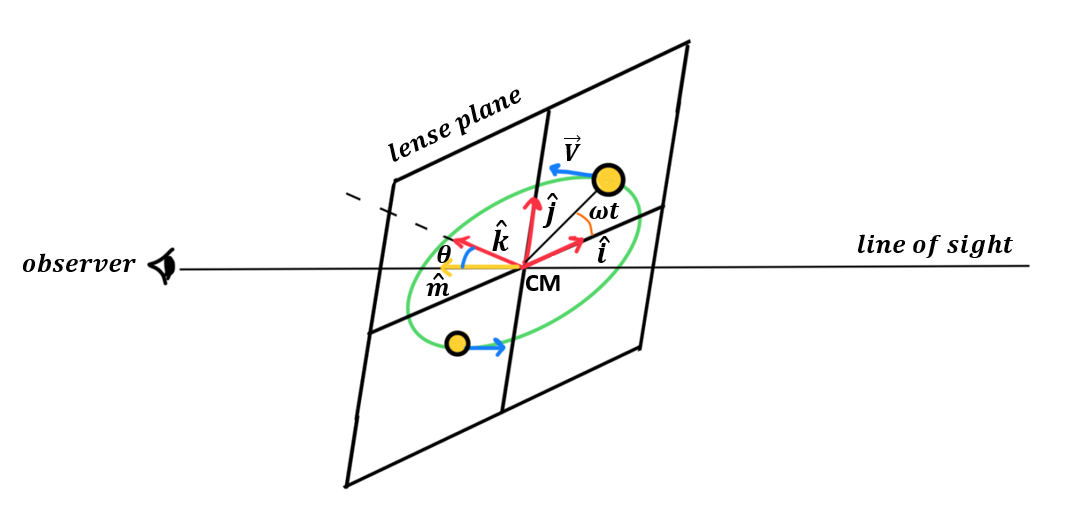}
\end{center}
\caption{ orbital velocity in binary lensing system
\label{fig:23}
}
\end{figure}

For a binary lens system, there is also an orbital velocity. The rotation velocity in the binary system can be considered as two horizontal and vertical components in the lens plane. equations as a correction.
\begin{equation}\label{equ:83}
\vec{V} = V\cos\omega t\hat{i}+ V\sin\omega t \ \hat{j}
\end{equation}
By considering a base vector along the line of site, we can decompose this rotation speed into components parallel to the line of site and perpendicular to it. 
\begin{equation}\label{equ:84}
\hat{m} = \sin\theta  \hat{i}+ \cos\theta \hat{k},
\end{equation}
Where $\theta$ is the angle between the  base vector along the perpendicular line to the lenses plane and the line of sight (see Figure \ref{fig:23} ) .In gravitational microlensing, only the component perpendicular to the line of sight is effective in the frequency shift, and this value can be entered into the equations as a correction.
\begin{equation}\label{equ:85}
\vec{V}.\hat{m} = \sin\theta \cos\omega t 
\end{equation}
\begin{equation}\label{equ:86}
\vec{V}-\vec{V}.\hat{m}=V\cos\omega t\hat{i}+ V\sin\omega t\hat{j}- \sin\theta \cos\omega t [ \sin\theta  \hat{i}+ \cos\theta \hat{k}]
\end{equation}
\begin{equation}\label{equ:87}
\vec{V}-\vec{V}.\hat{m}=(V\cos\omega t- \sin^2\theta \cos\omega t)\hat{i}+ V\sin\omega t\hat{j} - \sin\theta \cos\omega t \cos\theta \hat{k}
\end{equation}
This velocity component obtained here can be considered as a correction term for the entire distance of the lens relative to the observer.



\bsp	
\label{lastpage}
\end{document}